\pdfoutput=1

\documentclass[letterpaper,12pt]{scrartcl} 
\usepackage{subcaption}
\usepackage[american]{babel}
\usepackage[pdftex]{graphicx}
\usepackage{setspace} 
\usepackage{comment}
\usepackage{natbib}
\usepackage{amsmath, bm}
\usepackage[utf8]{inputenc}
\usepackage{booktabs} 
\usepackage{threeparttable}
\usepackage{varioref}
\usepackage{tikz}
\usepackage[final]{pdfpages}
\usepackage{multirow}
\usepackage{amssymb}
\usepackage{ragged2e}
\usepackage{lscape}
\usepackage{url}
\usepackage{rotating} 
\usepackage{amssymb}
\usepackage{mathptmx}
\usepackage{url}
\usepackage{tikz}
\usepackage{color} 
\usepackage{times}
\usepackage{palatino}
\fontencoding{T1}
\usepackage[T1]{fontenc}
\usepackage[sc]{mathpazo}
\usepackage{titletoc}
\usepackage{enumitem}
\usepackage{authblk}
\usepackage{tabularx}

\newcommand{\indep}{\rotatebox[origin=c]{90}{$\models$}}

\usepackage{sectsty}
\setcapindent{0em}
\setkomafont{captionlabel}{\bfseries \normalsize} 
\setkomafont{caption}{\normalsize}
\usepackage{geometry}
\bibpunct[: ]{(}{)}{;}{a}{}{,~}

\setkomafont{sectioning}{\rmfamily\bfseries}
 \makeatletter
 \renewcommand\@makefntext[1]{%
 \parindent 1em%
 \textsuperscript{\@thefnmark}~#1%
 \vskip 0.2cm}
 \makeatother

\usepackage[hang]{footmisc}

\title{\normalsize{Leveraging the Power of Place: A Data-Driven Decision Helper to Improve the Location Decisions of Economic Immigrants}}

\author[1,2,$\ast$]{\small Jeremy Ferwerda}
\author[1,$\ast$]{ \small Nicholas Adams-Cohen}
\author[1,3,$\ast$]{ \small Kirk Bansak}
\author[1]{\small Jennifer Fei} 
\author[1]{\small Duncan Lawrence}
\author[1,4]{\small Jeremy Weinstein}
\author[1,4,5,$\dagger$]{\small Jens Hainmueller \vspace{-.5cm}}

\affil[1]{\scriptsize Immigration Policy Lab, Stanford University}
\affil[2]{Department of Government, Dartmouth College}
\affil[3]{Department of Political Science, University of California San Diego}
\affil[4]{Department of Political Science, Stanford University}
\affil[5]{Graduate School of Business, Stanford University}
\affil[$\ast$]{Equal contributor} 
\affil[$\dagger$]{Project director and corresponding author. Contact: jhain@stanford.edu. \vspace{-.5cm}}

\date{\small July 2020 \vspace{-1cm} }

\begin{document}

\pagenumbering{gobble}

\begin{figure}
\vspace{-2.3cm}
\hspace{-2.7cm}
\includegraphics[width=1\paperwidth]{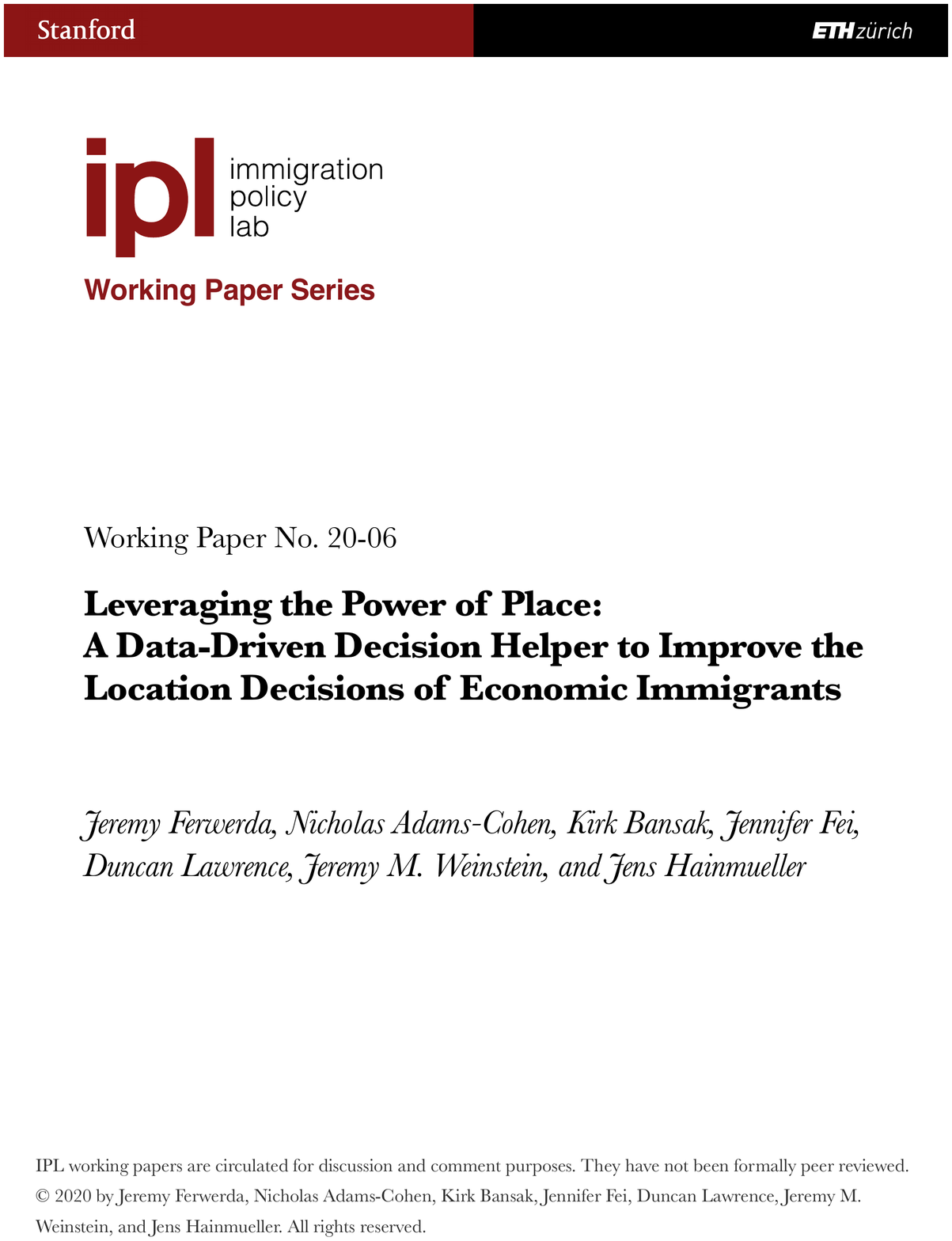}
\end{figure}

\maketitle
\pagenumbering{arabic}
\setcounter{page}{0}

\thispagestyle{empty}

\begin{abstract}

{\small

\noindent A growing number of countries have established programs to attract immigrants who can contribute to their economy. Research suggests that an immigrant's initial arrival location plays a key role in shaping their economic success. Yet immigrants currently lack access to personalized information that would help them identify optimal destinations. Instead, they often rely on availability heuristics, which can lead to the selection of sub-optimal landing locations, lower earnings, elevated outmigration rates, and concentration in the most well-known locations. To address this issue and counteract the effects of cognitive biases and limited information, we propose a data-driven decision helper that draws on behavioral insights, administrative data, and machine learning methods to inform immigrants' location decisions. The decision helper provides personalized location recommendations that reflect immigrants' preferences as well as data-driven predictions of the locations where they maximize their expected earnings given their profile. We illustrate the potential impact of our approach using backtests conducted with administrative data that links landing data of recent economic immigrants from Canada's Express Entry system with their earnings retrieved from tax records. Simulations across various scenarios suggest that providing location recommendations to incoming economic immigrants can increase their initial earnings and lead to a mild shift away from the most populous landing destinations. Our approach can be implemented within existing institutional structures at minimal cost, and offers governments an opportunity to harness their administrative data to improve outcomes for economic immigrants.

}
\end{abstract}

\clearpage

\section{Introduction}

Immigration has long been recognized as a driver of economic growth \citep{peri2015stem,kerr2016global}. Immigrants increase the size and diversity of the workforce, fill skill shortages, start businesses, and contribute to innovation \citep{hunt2010much,borjas1995economic,burchardi2020immigration}. To encourage these positive effects, many countries have complemented family-based and humanitarian admission streams with economic immigration programs, which prioritize the admission of skilled professionals. A prominent example is Canada’s Express Entry system, through which approximately 100,000 immigrants are admitted each year. Applicants earn points for qualifications, such as language ability, educational degrees, and occupational experience, as well as other factors that have been shown to be associated with long-term economic success in Canada. Applicants who are above a certain threshold for that particular application round receive invitations to apply for permanent residence \citep{Desiderio_2016}. Several other countries, such as Australia, New Zealand, and the United Kingdom, have implemented similar policies \citep{kerr2017high}.

The goal of these programs is to admit immigrants who are likely to succeed economically and contribute to the destination country. Yet despite these programs' intentions, immigrants nevertheless face a number of barriers to economic success. For instance, while a subset of individuals will have a preexisting job offer, the majority must select an initial location within the destination country in which to settle and begin their job search. However, immigrants generally lack access to personalized information on the locations that are aligned with their preferences and skill sets. As a result, the initial location decision can be affected by a variety of decision-making biases, such as availability heuristics. When economic immigrants choose suboptimal landing locations, economic admission programs cannot realize their full potential. Indeed, previous studies have demonstrated that the initial location of immigrants has a sizable impact on their short- as well as long-term economic outcomes \citep{aaslund2007and,damm2014neighborhood,bansak2018improving}. 

In this study, we propose a data-driven decision helper that leverages administrative data and machine learning methods to improve the initial location decisions of economic immigrants. Drawing on behavioral insights that have been used to improve decisions in other policy domains \citep{thaler2009nudge,oecd2017behavioural}, our approach seeks to enhance the choice architecture that shapes immigrants' decisions by providing them with systematic personalized information, delivered in the form of informational nudges, about which locations in the destination country would likely be beneficial to them. Building upon the outcome-based matching algorithm developed in \citet{bansak2018improving}, the decision helper provides newly invited immigrants with location recommendations that align with their preferences and maximize their expected economic outcomes. These recommendations draw on machine learning models applied to administrative data, which predict how immigrants with similar profiles have fared across possible landing locations, in combination with elicited preferences. The recommendations from the decision helper are not meant to be binding, but provide additional information that assists newly invited economic immigrants to make more informed location decisions.

To illustrate the potential of our approach, we evaluate administrative tax and landing data on recent cohorts from the economic immigration programs within Canada's Express Entry system. We find that landing locations are highly concentrated, and many economic immigrants settle in destinations that are sub-optimal from the perspective of predicted earnings. Using backtests and simulations, we find that providing data-driven location recommendations could significantly increase the annual income of economic immigrants and more widely disperse the benefits of economic migration across Canada. These gains would be realized at limited marginal cost, since the Canadian government already collects the administrative data used to train the models and communicates regularly with economic immigrants throughout the application process. Although the decision helper should be tested prospectively via a randomized controlled trial to evaluate its full impact on a variety of outcomes, our results suggest that nudging incoming economic immigrants with personalized information could improve their outcomes and create an opportunity for governments to leverage their existing data to offer an innovative resource at scale. 

\section{A Data-Driven Decision Helper}

\subsection{Motivation} 

Our approach is motivated by a growing body of research that demonstrates the importance of the initial landing location in shaping immigrants' outcomes. For example, studies have used quasi-experimental designs to demonstrate that an immigrant's initial landing location has an impact on short- as well as long-term economic success \citep{aaslund2007and,damm2014neighborhood,bansak2018improving}. Similarly, when examining outcomes among non-immigrants, experiments have shown that families who were randomly offered housing vouchers to move to lower-poverty neighborhoods had improved long-term outcomes in terms of earnings and educational attainment for their children \citep{chetty2016effects,ludwig2013long}. 

The initial destination choice is also consequential given that many immigrants tend to remain in their landing location \citep{kaida2020refugees,mossad2019search}. For example, in Canada, more than 80\% of recent economic class immigrants remained in their arrival cities ten years later \citep{kaida2020refugees}. In addition, landing locations are often highly concentrated. For example, in Australia, more than half of all recent immigrants settled in Greater Sydney and Greater Melbourne, and only 14\% settled outside the major capital cities \citep{Sajeda2019}. If initial settlement patterns concentrate immigrants in a few prominent landing regions, many areas of the country may not experience the economic growth associated with immigration. Moreover, undue concentration may impose costs in the form of congestion in local services, housing, and labor markets. To address the uneven distribution of immigrants, governments including Canada and Australia have implemented policy reforms to regionalize immigration and encourage settlement outside of well-known major cities \citep{taylor2014benefits,fotros2018destination,hugo2008immigrant,brezzi2010determinants}.

Although the evidence suggests that the initial landing location shapes immigrants' outcomes, choosing an optimal destination from the large set of potential options is a formidable task. Research suggests that immigrants consider the location of family or friends, the perceived availability of employment opportunities, or preferences regarding climate, city size, and cultural diversity \citep{chiswick2004immigrants,hyndman2006size,akbari2007initial,massey2008new,brezzi2010determinants,tonkin1993initial,damm2009determinants,mossad2019search}. While some of these considerations may lead immigrants to correctly identify an optimal location, research also suggests that location decisions are impaired by common cognitive biases. One such bias, which has been well documented as a powerful influence across many choice settings with incomplete information \citep{tversky1974judgment,thaler2009nudge}, is an availability heuristic. This heuristic suggests that immigrants prioritize places they have heard about, and that they overlook less prominent locations even though these locations may actually align with their preferences and skills.

For example, in the Canadian context, studies indicate that many location decisions are linked to the international prominence of destinations \citep{di2005immigrant}. As \citet{begin2010immigrant} argues, ``many immigrants choose Toronto simply because that is all they know of Canada'' (also see \citet{mcdonald2004toronto}). Similarly, \citet{teo2003imagining} concludes that ``unfamiliarity means that decisions regarding their initial destination are often reliant on secondary information sourced from earlier migrants, immigration companies, the Internet or other sources'' (also see \citet{fotros2018destination}). Recognizing that perceptions of places can skew location decisions toward prominent cities, some provinces have attempted to influence perceptions by providing prospective immigrants with information about less prominent locations. Evaluating these programs, \citet{begin2010immigrant} notes that when ``prospective immigrants are provided with more information and provided with more choices, they often choose differently.'' 

Interventions that use behavioral insights to counteract the effects of cognitive biases have been used in a wide variety of policy domains \citep{thaler2009nudge,oecd2017behavioural}. Our approach builds on these behavioral insights, and seeks to enhance the choice architecture for newly invited economic immigrants by systematically providing personalized recommendations as they make a decision about where to settle in the destination country. The recommendations from our decision helper reflect individuals' location preferences as well as data-driven predictions about the locations where they are likely to attain the highest earnings given their profile. The recommendations thus act as informational nudges that counteract the effects of limited information and cognitive biases, assisting immigrants in making more informed location decisions.

The primary anticipated users of the tool are economic immigrants who have been invited to apply for permanent residence via an economic admissions stream and are in the process of selecting a landing location within the destination country. Since governments communicate regularly with immigrants during this transition period to provide information about the immigration process, the decision helper could be offered online via a user interface at minimal cost. We expect that immigrants' likelihood of using the decision helper tool will depend on their prior level of certainty in their destination decision. As a result, the decision helper provides a complementary source of information with minimal disruption to existing sets of resources that guide decision making. Our decision helper could---with appropriate adjustments---also be useful for other immigrants or even Canadian-born residents as an informational tool in deciding if and where to move. Note that even Canadian-born residents typically do not have access to granular administrative data which would allow them to discern how workers with similar skill-sets and backgrounds fare in various locations.   

\subsection{Design}

\begin{figure}[tbhp]
\centering
\includegraphics[width=1\linewidth]{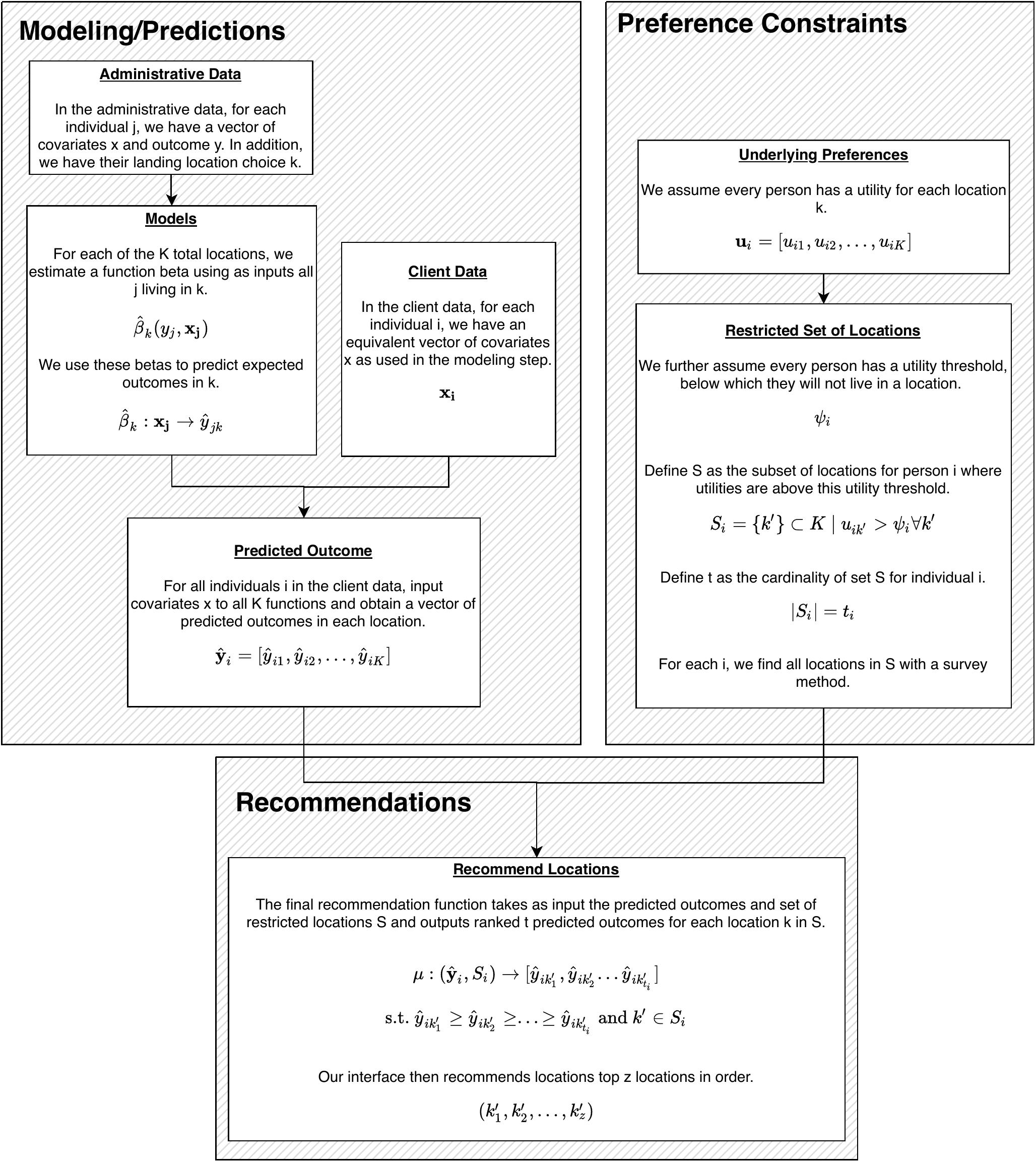}
\caption{Decision Helper}
\label{fig:decisionhelper} 
\end{figure}

The decision helper approach combines three main stages: modeling and prediction, preference constraints, and recommendations. Figure \ref{fig:decisionhelper} is a flowchart of these different steps. 

\subsubsection{Modeling/Prediction} 

Our approach leverages individual-level administrative data from prior immigrant arrivals. Governments routinely gather information on applicants to economic immigration programs, ranging from individuals' skills, education, and prior job experience to their age, gender, and national origin. Using unique identifiers, these background characteristics can be merged with applicants' initial landing locations and economic outcomes, such as earnings or employment. Although governments with economic admission streams collect these data as part of normal program administration, to our knowledge they have not been systematically leveraged to predict how immigrants with different profiles fare across various landing locations.

Our approach leverages these historical data to fit a set of location-specific, supervised machine learning models that serve as the basis for recommended landing locations for future immigrants. These models learn how immigrants' background characteristics and skill sets are related to taxable earnings within each potential landing location, while also accounting for local trends over time. The models can then be used to predict an economic immigrant's expected earnings at any of the possible landing locations. To minimize bias in the models' earnings predictions, a broad set of characteristics related to immigrants' backgrounds, qualifications, and skills are included as predictors. Furthermore, to reduce the possibility that observed patterns are driven by location-specific self-selection bias, the data used to train the model could exclude prior immigrants with standing job offers, family ties to specific locations, or other special situations. See the Supplementary Material (SM) appendix for a formalization and decomposition of the possible selection bias in the models, along with discussion on how such bias can be limited.

\subsubsection{Preference Constraints} 

Next, the decision helper elicits and incorporates individuals' location preferences. Even if a particular location is predicted to be the best for an immigrant in terms of expected earnings, if the immigrant strongly prefers not to live there, then recommending such a location will result in limited uptake and downstream dissatisfaction. To accommodate individual preferences, we identify the list of locations each user would consider unacceptable, and then limit the choice set to the remaining locations prior to optimizing on expected earnings. 

Our approach is agnostic to the specific method used to rule out unacceptable locations prior to making recommendations. For instance, users can be directly queried to indicate regions of the country where they are unwilling to settle. Alternately, users can be asked to provide preferences regarding specific location characteristics via direct questioning or conjoint survey tasks, spanning identifiable features such as urban density, climate, and the relative availability of amenities. These can be mapped onto observable characteristics of each landing location in order to identify acceptable locations according to the expressed preferences.

\subsubsection{Recommendations} 

After taking individual preferences into account to constrain the choice set, the remaining locations are then ranked with respect to the individual's predicted earnings in each location. The decision helper delivers these recommendations to the user in the form of an informational nudge via an online interface. Users can be given either a reduced number of the top ranked locations (e.g. the top 3 locations) or a full ranked list of the locations, along with accompanying information. Users choose whether to use the tool and follow the recommendations. This approach is non-coercive, seeking only to inform those who can benefit while not interfering with immigrants who may already have solid plans or private information guiding their selection of a particular location.

\section{Empirical Analysis}

We illustrate the potential of a decision helper to recommend initial landing locations for newly invited immigrants by evaluating data from Canada's flagship Express Entry system. Express Entry is a system that manages applications to Canada’s high skilled economic immigration programs, which select skilled workers for admission through a points-based system.\footnote{These economic immigration programs are the Federal Skilled Worker Program, the Federal Skilled Trades Program, the Canadian Experience Class, and a portion of the Provincial Nominee Program.} The Express Entry application process involves several stages and is designed to select applicants who are most likely to succeed in Canada \citep{immigration2019express}. First, eligible candidates create and submit a profile to indicate their interest in moving to Canada. If candidates meet the minimum requirements for one of the programs managed by Express Entry, they are entered into the Express Entry pool, awarded points based on information in their profile and ranked according to the Comprehensive Ranking System (CRS). The CRS awards points based on human capital characteristics including language skills, education, work experience, age, employment and other aspects that previously have been shown to be associated with long-term economic success in Canada. Factors in the CRS are generally grouped under two categories: core points and additional/bonus points. Candidates with the highest rankings in the pool are invited to apply online for permanent residence following regular invitation rounds. If candidates' CRS scores are above a specified threshold for an invitation round, they receive an invitation to apply for permanent residence, to be submitted within 90 days of receiving an invitation \citep{immigration2019express}. If an application is approved by an IRCC officer, permanent resident visas are issued so that the applicant and his or her accompanying family members can be admitted to Canada. Processing times for Express Entry profiles vary depending on the program of admission, but the majority of applications are processed within six months.  

Since the system initially launched in 2015, it has steadily expanded. In 2018, 280,000 Express Entry profiles were submitted, and 92,331 people were admitted to Canada through the Express Entry system \citep{immigration2019express}. The growing importance of this system is mirrored by similar developments within other advanced economies. For instance, Australia and New Zealand also use a similar expression of interest process to determine applicants' eligibility and offer invitations to apply for permanent residence. Countries such as Austria, Japan, South Korea, and the United Kingdom also have aspects of points-based admissions systems built into their economic immigration programs.     
    
\subsection{Data}

We draw on data from the Longitudinal Immigration Database (IMDB)---the integrated administrative database that Immigration Refugees and Citizenship Canada reports on the outcomes of immigrants. IMDB was initially a basic linkage between tax files and the Permanent Residents Database. The IMDB (2019 release) includes more than 12 million immigrants who landed in Canada between 1952 and 2018 and income tax records from 1982 to 2017, as well as all temporary residents, Express Entry Comprehensive Ranking System scores, citizenship uptake, and service usage for settlement programs. We subset the data to include principal applicants who arrived between 2012 and 2017 under the following programs: the Federal Skilled Worker program, the Federal Skilled Trades program, and the Canadian Experience Class. Applications for these admission streams are managed by the Express Entry system. We further subset the data to exclude individuals who were minors at the time of arrival ($\leq$18), as well as individuals who did not file a tax return while living in Canada. Given that the Express Entry system does not apply to Quebec, we also exclude all immigrants who first landed in Quebec or entered on an immigration program run by Quebec. The final sample size consists of 203,290 unique principal applicants. 
    
\subsection{Measures}
    
The outcome measure is immigrants' individual annual employment income, measured at the close of the first full calendar year after arrival. We model this outcome as a function of a variety of predictors that are either prior to or contemporaneous with an immigrant's arrival. Predictors used in the modeling stage include age at arrival, citizenship, continent of birth, education, family status, gender, intended occupation, skill level, English ability, French ability, having a prior temporary residence permit for study in Canada, having a prior temporary residence permit for work in Canada, having previously filed taxes in Canada, arrival month, arrival year, immigration category, and Express Entry indicator. See Table \ref{tab:descrstats} in the SM appendix for more information and summary statistics on these measures.
    
We map immigrants' landing locations to a specific Economic Region (ER) within Canada, using census subdivision codes. As regional predictors, we also include the population and the unemployment rate within the ER in the quarter of immigrants' arrival. An ER is a Canadian census designation that groups neighboring census divisions to proxy regional economies. We use the ER as the primary unit throughout the analysis. There are 76 ERs in total, but in our analysis there are 52 after excluding Quebec and merging the smallest ERs using standard census practices.
    
\subsection{Models}
    
The modeling approach is based on the methodology developed in \citet{bansak2018improving}. We first merge historical data on immigrants' background characteristics, economic outcomes, and geographic locations. Using supervised machine learning methods, we fit separate models across each ER estimating an immigrant's annual employment income as a function of the predictors described above. These models serve as the basis for the decision helper tool's recommendations, as they allow for the generation of annual employment income predictions across each ER that are personalized to each immigrant's background characteristics.
    
As our modeling technique, we use stochastic gradient boosted trees. We use 10-fold cross-validation within the training data to select tuning parameter values, including the interaction depth, bag fraction, learning rate, and number of boosting iterations.\footnote{The cross-validated R$^{2}$ for our primary set of models (where the units of analysis are principal applicants) is 0.54. Within the context of incomes -- which are highly skewed and difficult to predict -- this is relatively high. This represents a substantial improvement over the R$^{2}$ for an analogous linear regression model using cross-validation (0.34). See the SM appendix for more details, including a breakdown of the relative importance of the predictors in the boosted trees models.} More details are provided in the SM appendix.
    
\subsection{Simulations}
    
To estimate how our proposed decision helper would affect income and influence location decisions, we perform a series of backtests using historic Express Entry cohorts. Specifically, we implement a series of simulations in which the decision helper provides recommendations to individual immigrants. We then simulate uptake of these recommendations, and for individuals who follow the recommendation, we compare the expected income at that location and at the location where they actually landed.
    
After training the models, we input the background characteristics of 2015 and 2016 Express Entry principal applicants ($n=17,640$) to obtain predicted income across ERs. These predictions serve as the basis for the simulated recommendations. The degree to which immigrants would follow such recommendations is unknown. To model these dynamics, the simulations vary two parameters that reflect different assumptions concerning the influence of the recommendation on immigrants' location decisions. 
    
The first parameter is the compliance rate, denoted by $\pi$, which is defined as the probability that individuals will follow the recommendations. We assume that the probability of following a recommendation decreases linearly across income quantiles. We apply an upper bound $\pi_{max}$ to the individuals with the lowest actual income. We then linearly interpolate to a value of $\pi = 0$ across the income distribution. Each individual within the prediction set thus receives an individual compliance parameter, $\pi_{i}$. Functionally, the average compliance rate across the distribution is $\pi_{max}$/2. For example, in the simulations with $\pi_{max}=.30$, on average 15 percent of immigrants are expected to follow the recommendation, and the probability varies from a high of 30 percent for the lowest-income immigrants to close to zero percent for the highest-income immigrants. We chose to vary $\pi_{i}$ as a function of income under the assumption that wealthier individuals are more likely to have self-selected into location-specific employment opportunities. Simulations that impose a uniform compliance parameter instead also suggest substantial gains (see SM appendix).
    
The second parameter is the number of acceptable locations, denoted by $\phi$. Each applicant is assumed to have a set of idiosyncratic preferences regarding locations, which results in a ranked preference order of ERs, ranging from the most attractive (1) to the least attractive (52). The $\phi$ parameter determines how many top-preference-ranked ERs are included within the optimization. Location preferences are unmeasured within the administrative data, and must be inferred. Using the landing ER as the dependent variable, we fit a multinomial logit model and proxy immigrants' preferences using their predicted probability of landing in each ER. After obtaining predictions for each Express Entry case, we rank order locations by each individual's predicted probability of landing, randomly breaking ties. For each individual, the resulting preference ranks are then used in conjunction with the parameter $\phi$ to define their set of acceptable locations, which serves as the initial set of locations considered when selecting the locations with the highest expected incomes. To guarantee that gains are not entirely driven by subsets of locations with certain characteristics, we run simulations entirely removing certain ERs from consideration, and find no major deviation from our core results (see SM appendix).
    
We conduct a simulation for various combinations of parameters. For each immigrant we consider only the top $\phi$ preference-ranked locations, and return the three locations within this subset that are expected to yield the highest employment income for the immigrant. We assume that individuals who follow the recommendations have an equal probability of selecting each of these three locations. However, with probability (1-$\pi_{i}$), individuals will select their original location rather than any of the recommendations. For each case, we draw from a uniform distribution bounded by 0 and 1 to determine whether the case will take the recommendation or not. If (draw) > $\pi$, cases are assumed not to have followed the recommendation, and their location is recorded as their actual location. Their income is recorded as their predicted income within their actual location; for these immigrants there is no gain in income from using the tool. For immigrants where (draw) < $\pi$, we perform a second draw to determine which of the three locations they will select. For these immigrants, the expected gain in income is computed as the difference between the expected income they would earn in the recommended location and the location they would have chosen without the tool. After performing these random draws for each of the 2015 and 2016 Express Entry users, we obtain the total expected difference in location counts and income.

\section{Results from Empirical Analysis}

\subsection{Current Settlement Patterns}

\begin{figure}[tbhp]
\centering
\includegraphics[width=1\linewidth]{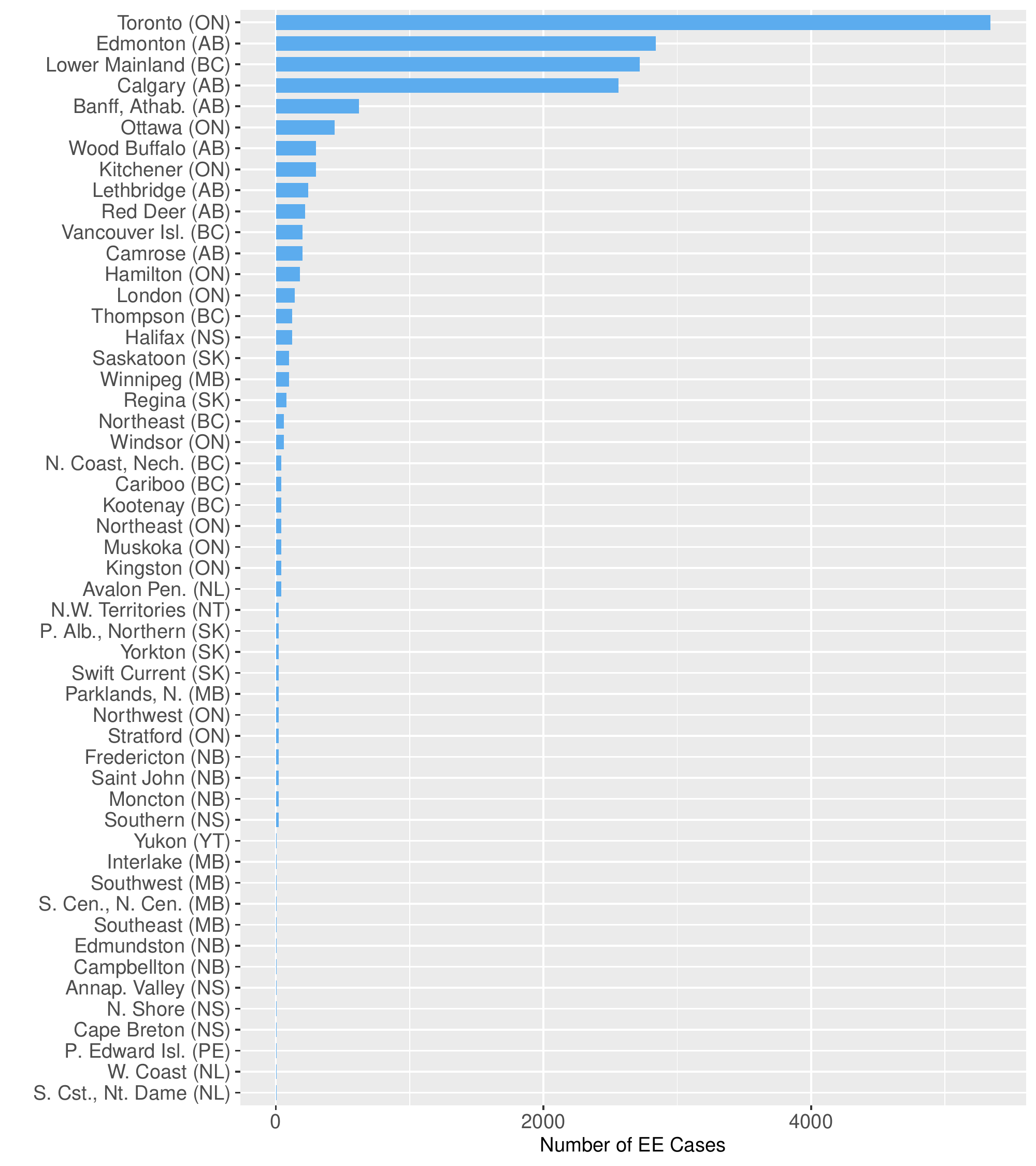}
\caption{Landing Economic Regions for Express Entry Cohorts Arriving 2015-2016. N=17,640.}
\label{fig:concentration}
\end{figure}

As shown in Figure \ref{fig:concentration}, economic immigrants who arrived in the 2015 and 2016 arrival cohorts through the Express Entry system are highly concentrated within a few regions of Canada. About 78\% settled in one of the four largest ERs as their initial destination, and 31\% of immigrants selected Toronto. In stark contrast, only about 44\% of the overall population is concentrated in those four ERs.\footnote{We exclude Quebec from this computation to have an accurate comparison.}

\begin{figure}[tbhp]
\centering
\includegraphics[width=1\linewidth]{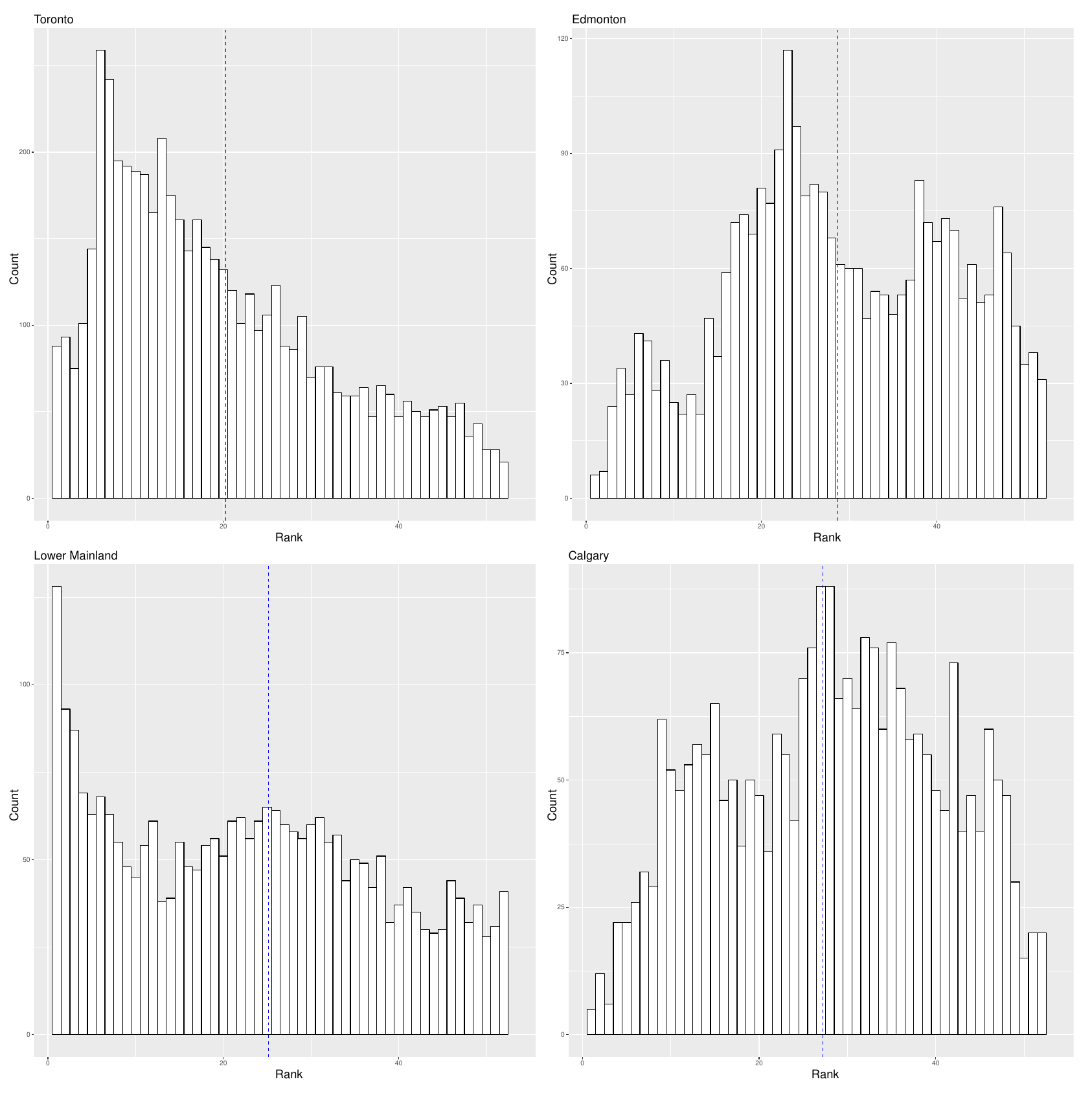}
\caption{Estimated Annual Income: Rank of Landing Economic Region vs. All Economic Regions. N=17,640}
\label{fig:ranks}
\end{figure}

To what extent do these concentrated settlement patterns support the goal of maximizing incomes? We evaluate this by estimating each economic immigrant's expected income in every potential landing region as a function of their background characteristics and qualifications (see SM for details). Among economic immigrants who selected one of the four most selected locations, Figure \ref{fig:ranks} displays how the selected ER would rank in terms of expected earnings relative to all other potential ERs. For example, a rank of 1 for a given immigrant in the top left panel indicates that, at the time of arrival, the models estimate that Toronto ranked first (i.e. best) out of 52 possible landing locations in terms of the expected employment income for that immigrant. 

Although the models suggest that a small subset of individuals selected an initial location that maximizes their expected income, we find that for many economic immigrants the chosen location is far from optimal in terms of expected income. For instance, among economic immigrants who chose to settle in Toronto, that landing location only ranked approximately 20th on average out of the 52 ERs in terms of maximizing expected income in the year after arrival. In other words, the data suggest that for the average economic immigrant who settled in Toronto, there were 19 other ERs where that immigrant had a higher expected income than in Toronto. The situation is similar for other prominent locations, including Edmonton, Lower Mainland, and Calgary, where the average ranks are 28, 24, and 26, respectively. Across Canada as a whole, the average rank is a mere 26.5. This suggests that many immigrants do not select locations where individuals with similar background characteristics tend to achieve the best economic outcomes, and there is potential to improve immigrants' landing choices. 

\subsection{Changes in Expected Income and Arrival Locations}

\begin{figure}[tbhp]
\centering
\includegraphics[width=1\linewidth]{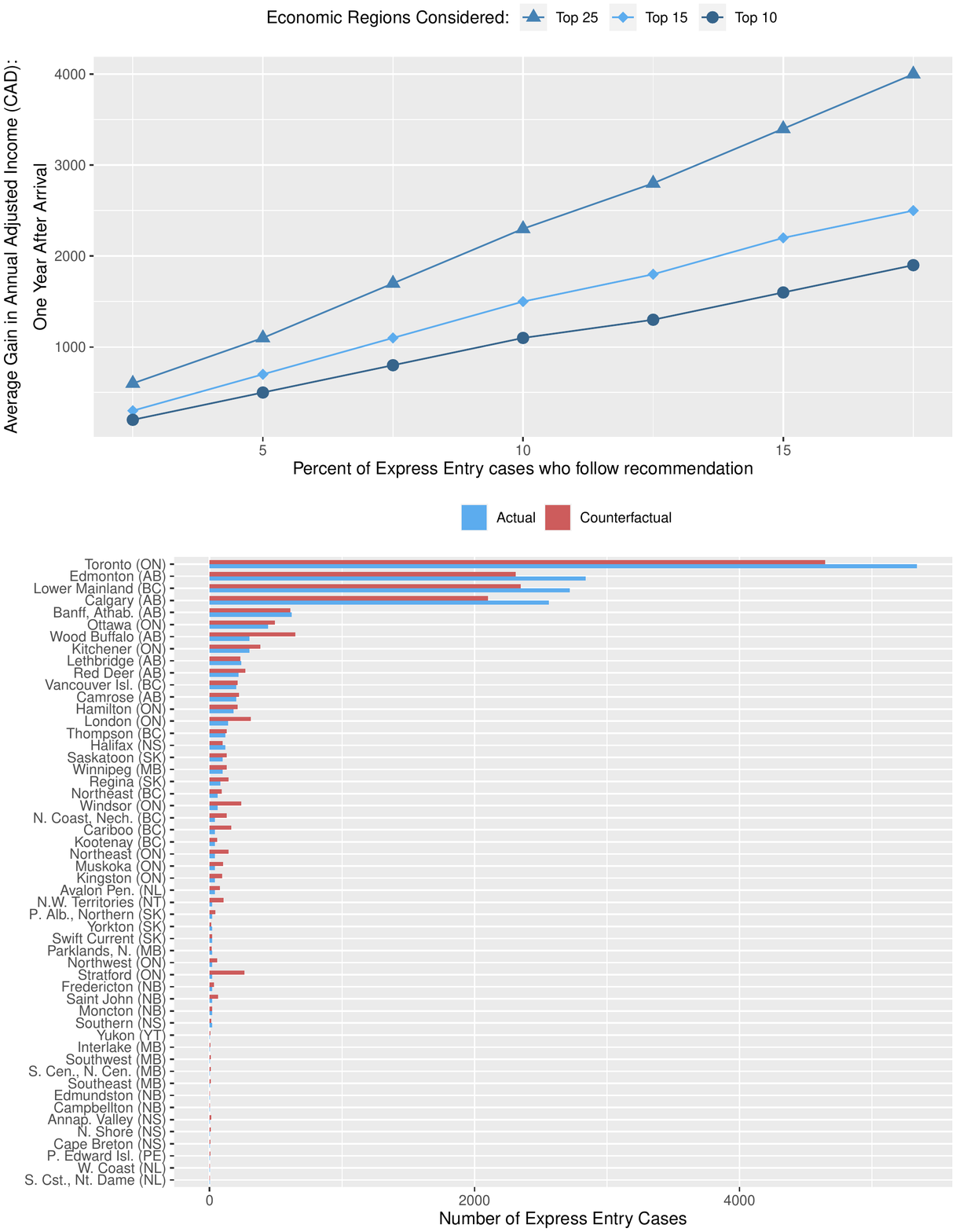}
\caption{Estimated Average Income Gains and Shifts in Arrival Locations. N=17,640}
\label{fig:gainsandmove}
\end{figure}

Figure \ref{fig:gainsandmove} displays the results from backtests that examine how a data-driven decision helper tool may influence the expected incomes and location decisions of economic immigrants who enter through the Express Entry system (see SM for details). The top panel shows the estimated effects on the average expected income one year after arrival for economic immigrants across the entire backtest cohort of 2015 and 2016 arrivals. We simulate effects using a varying set of parameters, including the share of immigrants who are assumed to follow the recommendations (horizontal axis) and the number of locations that are considered for each recommendation, based on the immigrants' modeled location preferences (colors and symbols). These results report the average gain from 100 simulation runs. 

The simulations suggest gains in expected annual incomes, even under scenarios in which compliance is low and/or location preferences are highly restrictive. For example, using the assumption that on average only 10\% of immigrants settle in one of the recommended locations, and that individuals’ location preferences will rule out 42 of the 52 possible locations as unacceptable (the scenario labeled ``Top 10''), the simulation yields an average gain in expected annual employment income one year after arrival of \$1,100, averaged across the full cohort. This amounts to a cumulative gain of \$55 million in total income for every 50,000 cases that enter Canada via the Express Entry system. Note that these gains are entirely driven by the 10\% of immigrants who follow the recommendations, since we assume zero gains for the rest of the cohort. Immigrants who do follow the recommendation increase their expected annual employment income one year after arrival by \$10,600 on average, relative to the estimated income at the location they would have selected without using the decision helper. These gains are large relative to the observed average first-year income within the prediction sample (\$49,900).

Across the full cohort, the total expected gains from implementing the decision helper would be larger if immigrants had less restrictive location preferences and/or more immigrants followed the recommendations. For example, under the assumption that 15\% of immigrants follow the recommendations, and the recommended locations are chosen from a set of 25 acceptable locations, the average expected annual income one year after arrival across the cohort increases by \$3,400. The SM demonstrates that these results are similar across various robustness checks, including replicating the analysis with cost of living adjusted income (Figure \ref{fig:s2}), recommending locations to maximize the joint income of principal applicants and their spouses (Figure \ref{fig:s3}), or removing smaller, larger, or growing ERs from consideration (Figure \ref{fig:s6}).

The lower panel in Figure \ref{fig:gainsandmove} displays the anticipated impact on the distribution of economic immigrants across initial landing regions under the 15\% compliance and Top 25 location simulation scenario. The results suggests that we would see a mild shift from the most populous destinations toward mid-sized landing regions. For instance, about 15\% of immigrants who chose one of the four largest locations would have chosen an alternate location in Canada if they had followed the recommendation. Although there is not a marked redistribution of arrivals toward the smallest locations, the estimates for smaller locations are likely conservative given that the preferences for our simulation were derived from data on the existing residential patterns of immigrants across Canada. Figure \ref{fig:s4} in the SM shows the expected distribution if no location preferences were taken into account. While these scenarios find that the majority of outflows continue to be associated with the four largest ERs, we find more movement into a subset of the smaller locations when we do not restrict locations based on the inferred preferences.

\subsection{Changes in Expected Income for Subgroups}

\begin{figure}[tbhp]
\centering
\includegraphics[width=1\linewidth]{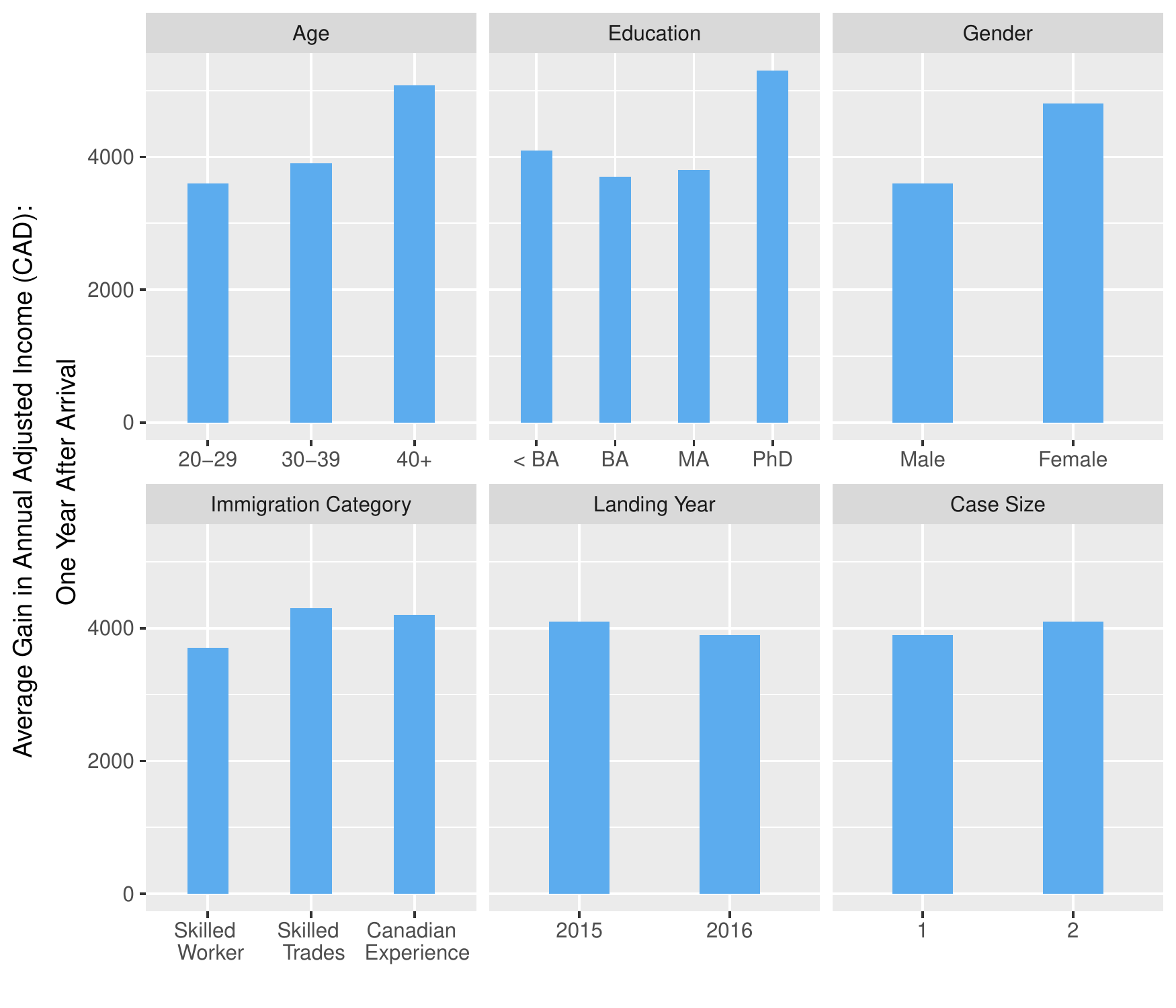}
\caption{Estimated Average Income Gains for Subgroups. N=17,640}
\label{fig:subgroups}
\end{figure}

We also assess the distribution of potential gains across subgroups to understand the differential effects our approach could have for economic immigrants of various backgrounds. Figure \ref{fig:subgroups} shows the estimates of the change in average expected incomes one year after arrival across a variety of different subgroups, again using the assumption that 15\% of immigrants would follow the recommendations and that recommended locations are chosen from a set of 25 acceptable locations. While expected gains vary as a function of individuals’ characteristics, the overall increase in income does not appear to be the result of disproportionate benefit on the part of any particular demographic or socioeconomic groups. Instead, we find comparable average gains across a range of subgroups of economic immigrants, including groups stratified by gender, education level, case size, landing year, and immigration category.\footnote{Results for case size subgroups are only shown for case sizes of 1 and 2 due to an insufficient number of cases of size greater than 2.}

\section{Potential Limitations}

The impact assessment is limited to backtests applied to historical data. Such backtests are commonly used to examine the potential impact of new approaches, but they cannot fully capture all the factors that may affect the potential impact of our decision helper in a prospective implementation. 

For instance, economic outcomes could be influenced by compositional effects if many immigrants with a similar profile receive the same location recommendation. Modeling such effects in a backtest is challenging because their direction and magnitude is theoretically ambiguous. An increased concentration could lower expected incomes due to saturation, or alternatively, increase incomes due to local agglomeration economies, which are common for high-skilled migration \citep{kerr2017high}. Similarly, economic immigrants account for a relatively small fraction of the local labor market, implying that the direct impact of the tool on labor market saturation is difficult to determine a priori. 

Although we do not model compositional effects directly in the backtests, in a prospective application the decision helper can learn potential compositional effects over time as the models are continually updated based on observed data from new arrivals, as well as local economic conditions at the time of arrival. Therefore, should a particular location cease to be a good match for a particular immigrant profile due to increased concentration, the models would adjust to this pattern over time and no longer recommend that location. In addition, the approach can be adjusted to incorporate location-specific quotas if desired by governments.

The results in a prospective implementation could also differ from the backtests if the expected incomes predicted by the machine learning models over- or under- estimate the actual incomes that newly invited immigrants attain if they were to follow the recommendations. Such prediction errors could occur for a variety of reasons, including immigrants selecting into locations based on unobserved characteristics. Such unobserved characteristics can be separated into two broad categories. The first category includes unobserved characteristics that are unrelated to any particular location, and hence could be associated with higher (or lower) earnings potential in all possible locations. Examples would include an individual's unmeasured abilities or motivation. The second category includes unobserved characteristics that are unique to specific locations, and hence represent an earnings advantage for a particular individual only in a select location (or select locations). Examples would include an individual's unknown job offer or social network (e.g. family members) in a specific location. See the SM appendix for a more comprehensive discussion, formalization, and decomposition of the possible selection bias.

These concerns about selection bias driving the results in our backtests are partially addressed by the fact that we train flexible models on a rich set of covariates derived from application data to economic admissions programs. That is, conditional on the broad set of background and skills-based characteristics we observe, it is not likely that individuals have self-selected into locations as a function of unobserved non-location-specific characteristics that account for the full magnitude of the estimated gains we observe in the backtests. For instance, for individuals who are identical on the observed characteristics (e.g. same age, education, profession, skills, etc.), their variation in unmeasured variables such as motivation would need to both be integral in their location choices and significantly affect their earnings potential (see SM for details).

Finally, while the backtest results suggest the possibility of gains if the decision helper were implemented, it is important to obtain reliable estimates of impact through a prospective randomized-controlled trial (RCT). An RCT would randomly assign new arrivals to receive recommendations from the decision helper, allowing for a rigorous evaluation of the tool's impact on a variety of outcomes, including incomes, satisfaction, and location patterns. 

\section{Potential Risks}

To examine potential risks, it is important to consider how introducing a decision-helper tool would influence the status quo. The decision-helper provides additional information to incoming economic immigrants so that they can make more informed location decisions, from among the set of possible locations that match their preferences. In doing so,  it does not limit immigrants' agency to choose their final settlement location. Given that the recommendations are based on historical data, the predictions may be subject to error. As a result, the decision helper should be viewed as a complement, rather than a replacement, of the existing information streams and processes that governments use to inform immigrants about potential destinations.

Care needs to be taken to transparently communicate the locational recommendations to users. In particular, users should be made aware that the recommendations are based upon the goal of maximizing the particular metric of near-term income (or whichever specific metric has been applied) and that the predictions reflect the outcomes that recent immigrants with similar profiles have attained in the past. This does not guarantee that the user's realized income will be optimal at the recommended location, as expected income cannot take into account all possible factors that are unique to an individual. Nor does it imply that the recommended location would necessarily be optimal in terms of other possible life goals or long-term earnings. While we do not have evidence that selecting locations based on near-term earnings will have a negative impact on these longer-term outcomes, this is a theoretically possible risk that would need to be monitored over the course of a prospective implementation. 

\section{Discussion}

A growing number of countries have implemented economic immigration programs to attract global talent and generate growth. However, economic immigrants lack access to personalized information that would help them identify their most beneficial initial settlement locations. In this study, we propose a data-driven decision helper that delivers informational nudges to counteract the effects of limited information and cognitive biases. The decision helper harnesses insights from administrative data to recommend the locations that would maximize their expected incomes and align with their preferences. We illustrate its potential by conducting backtests on historical data from the Canadian Express Entry system. The results suggest that economic immigrants currently select sub-optimal locations, and that there could be gains in expected incomes from providing data-driven, personalized location recommendations. While the results from our backtests suggest potential gains in the Canadian context, it is important to assess the impact in the context of a pilot initiative with a randomized controlled trial. 

The decision helper outlined in this study is adaptable and could be rapidly implemented within existing institutional structures in various countries. First, as we demonstrate via application to Canadian data, many governments are already collecting administrative data that can be used to generate recommendations. Moreover, governments administering economic admission programs engage in regular communication with applicants, implying that the decision helper can easily be made accessible to a large group of users. In light of the gains observed in the backtests, these limited costs suggest a positive return on investment, even in scenarios where only a small share of immigrants follow the recommendations. Second, the approach is flexible in terms of implementation and can be adjusted to the specific priorities identified by the destination government. For example, the decision helper could be used to improve other measurable integration outcomes (for example, longer-term measures of income), incorporate location-specific quotas, and allow for a wide variety of approaches to elicit preferences and display recommendations. Third, the approach is designed as a learning system such that the models for the predictions are continually updated using observed data from new arrivals and changing local economic conditions. The decision helper therefore learns synergies between personal characteristics and landing locations as they evolve over time and adjusts the recommendations accordingly. Finally, the approach supports individual agency. The decision helper provides immigrants with personalized recommendations that help them make a more informed decision, but immigrants decide whether to use it, and they can decline the recommendations. The decision helper thus complements rather than replaces existing information streams and processes.

In sum, a data-driven decision helper holds the potential to assist incoming economic immigrants in overcoming informational barriers and choosing better landing locations. In addition, the approach we outline offers governments the ability to leverage administrative data to increase economic returns within the structures of their existing admission process. Together, we expect these factors to improve the well-being of economic immigrants and the communities in which they settle.

\section{Acknowledgments}

\noindent This study was completed as part of a Data Partnership Arrangement between Immigration, Refugees and Citizenship Canada (IRCC) and the Immigration Policy Lab. The analysis, conclusions, opinions, and statements expressed in the material are those of the authors, and not necessarily those of the IRCC. This research received generous support from Eric and Wendy Schmidt by recommendation of Schmidt Futures. We also acknowledge funding from the Charles Koch Foundation. These funders had no role in the data collection, analysis, decision to publish, or preparation of the manuscript.

\bibliographystyle{apsr}

\bibliography{circulation}

\clearpage

\appendix

\begin{center}
\Huge{
\textbf{Supplementary Material}
}
\end{center}

\vspace{1cm}

\setcounter{table}{0}
\setcounter{figure}{0}
\setcounter{equation}{0}
\renewcommand{\thetable}{S\arabic{table}}
\renewcommand{\thefigure}{S\arabic{figure}}
\renewcommand{\theequation}{S\arabic{equation}}
\renewcommand{\thesection}{S\arabic{section}}
\renewcommand{\thesubsection}{S\arabic{section}.\arabic{subsection}}

\section{Decision Helper Workflow}

In this section, we provide additional details and formalize our decision helper approach. This workflow consists of three stages, modeling/prediction, preference constraints, and recommendations. We repeat the visualization of the workflow with additional formal notation in Figure \ref{fig:decisionhelper}, and describe each step of the process in detail.

\subsection{Modeling/prediction}

In the first stage, we use training data to build a series of models that predict expected outcomes in a particular location. This process begins by gathering a set of \textbf{Adminstrative Data}, an individual-level dataset containing information about prior immigrant arrivals. This dataset must consist of individuals that are similar to the eventual users of the decision helper tool. \\

For each individual in the administrative data, we need three pieces of information: a collection of covariates, a measurable outcome, and a choice of landing location. Because our goal is to determine unique synergies between individual-level profiles and outcomes in a particular landing spot, we estimate models for each location separately. \\

For each individual in the administrative data $j = 1, \dots, m$, let the outcome of interest be denoted $y_j$ and the landing decision denoted $w_j \in \{1,\dots, K\}$. Let $\vec{x}_j$ represent a $p$-dimensional vector of relevant covariates for individual $j$, and $x_{ir}$ represent the $r$-th feature in the $p$-dimensional vector. Our goal in the model training portion of the workflow is to predict the outcome based on the relevant covariates and specified landing location; that is, to estimate function $\beta$ mapping $\vec{x}_j$ to $y_j$. As we want to find separate functional forms for each location $k = 1, \dots, K$, we estimate K total functions $\beta_k(\vec{x}_j|w_j = k)$.  We find an  approximation $\hat{\beta}_k$ to $\beta_k$ by minimizing the expected value of a specified loss function $L(y_j, \beta_k(\vec{x}_j) $ over the joint distribution of $(y, \vec{x})$:

\[
\hat{\beta}_k =\underset{\beta_k}{\mathrm{argmin}} \: \mathbb{E}_{(y,\vec{x})} \: L(y, \beta_k(\vec{x}))
\]

After estimating $\hat{\beta}_k$ for all $K$ landing locations in the administrative data, we then apply these models to the \textbf{Client Data}, the potential users of the decision helper tool. For each client $i = 1,\dots, n$, we have the same set of covariates used to train the models $\vec{x}_i$. By inputting $\vec{x}_i$  to each of the $\hat{\beta}_k$ location functions, we obtain individual predicted outcomes in each possible location. The following delineates each step in this process:

\begin{enumerate}
    \item Denote the administrative data by matrix $\mathbf{A}$
    \[
\mathbf{A} = \begin{bmatrix} 
    y_{1} & w_1 &  x_{11} & \dots & x_{1r} & \dots & x_{1p}  \\
    \vdots & \vdots &  \vdots &\ddots & \vdots &\ddots & \vdots\\
    y_{j} & w_j &  x_{j1} & \dots & x_{jr} & \dots & x_{jp}  \\
    \vdots & \vdots &  \vdots & \ddots & \vdots &\ddots & \vdots\\
    y_{m} & w_m &  x_{m1} & \dots & x_{mr} & \dots & x_{mp}  \\

    \end{bmatrix}
\]

\item Train a set of K models, 
\[\mathbf{L} = \{\hat{\beta_1}(\vec{x}_j),\dots, \hat{\beta_k}(\vec{x}_j), \dots , \hat{\beta_K}(\vec{x}_j) \}\] 
as follows.

For $k = 1, \dots, K$:
\begin{enumerate} 
    \item Subset $\mathbf{A}$ to individuals for whom $w_j = k$ and call this $\mathbf{A}_k$

    \[
\mathbf{A}_k =\begin{bmatrix} 
    y_{1} & x_{11} & \dots & x_{1r} & \dots & x_{1p}  \\
    \vdots &  \vdots &\ddots & \vdots &\ddots & \vdots\\
    y_{j} &  x_{j1} & \dots & x_{jr} & \dots & x_{jp}  \\
    \vdots  &  \vdots & \ddots & \vdots &\ddots & \vdots\\
    y_{m_l}  &  x_{m_k1} & \dots & x_{m_kr} & \dots & x_{m_kp}  \\

    \end{bmatrix}_{w=k}
= 
\begin{bmatrix} 
    y_{1} & \vec{x}_{1}  \\
    \vdots &  \vdots \\
    y_{j} &  \vec{x}_{j}  \\
    \vdots  &  \vdots\\
    y_{m}  &  \vec{x}_{m_k}   \\

    \end{bmatrix}_{w=k}
\]

Where $m_k$ denotes the number of individuals in the administrative data for whom $w_j = k$.

\item Using the data in $\mathbf{A}_k$, model and estimate $\hat{\beta}_k$.

Note that while there are many ways to potentially model  $\hat{\beta}_k$, we have found that using supervised machine learning methods provides the best flexible solution to capture complex non-linearities, interactions between covariates, and automatically engage in feature selection. To avoid overfitting on the training set, wherin $\hat{\beta}_k$ has very high predictive power in the training set but low out-of-sample predictive power, it is necessary to use cross-validation in the training process. 
\end{enumerate}

\item Denote the client data by matrix \textbf{C}.

    \[
\mathbf{C} =\begin{bmatrix} 
     x_{11} & \dots & x_{1r} & \dots & x_{1p}  \\
     \vdots &\ddots & \vdots &\ddots & \vdots\\
    x_{j1} & \dots & x_{jr} & \dots & x_{jp}  \\
      \vdots & \ddots & \vdots &\ddots & \vdots\\
     x_{n1} & \dots & x_{nr} & \dots & x_{np}  \\

    \end{bmatrix}
= 
\begin{bmatrix} 
    \vec{x}_{1}  \\
      \vdots \\
     \vec{x}_{j}  \\
      \vdots\\
      \vec{x}_{n}   \\

    \end{bmatrix}
\]

\item For all clients in $\mathbf{C}$ and all $k$ locations, estimate $\beta_k : \vec{x}_i \to \hat{y_i}$ as follows:

For $i = 1,\dots,n$

$\:\:\:\:\:\:$ For d = 1, \dots , k

$\:\:\:\:\:\:$ $\:\:\:\:\:\:$ Estimate $\beta_k(\vec{x}_i)$ by applying the $k$-th model in $\mathbf{L}$ to $\vec{x}_i$, where $\hat{\beta}_k(\vec{x}_i) = \hat{y}_{ik}$

$\:\:\:\:\:\:$ Arrange  $\hat{y}_{ik}$ into a vector $ \vec{\hat{y}}_{i} = [\hat{y}_{i1}, \hat{y}_{i2}, \dots, \hat{y}_{iK} ] $ 

\item Produce a matrix of predicted outcomes $\mathbf{M}$, with rows corresponding to clients and columns responding to potential landing locations as follows.

    \[
\mathbf{M} = \begin{bmatrix} 
    \vec{\hat{y}}_{1}  \\
      \vdots \\
     \vec{\hat{y}}_{i}  \\
      \vdots\\
      \vec{\hat{y}}_{n}   \\

    \end{bmatrix}
    =
    \begin{bmatrix} 
     \hat{y}_{11} & \dots & \hat{y}_{1k} & \dots & \hat{y}_{1K}  \\
     \vdots &\ddots & \vdots &\ddots & \vdots\\
    \hat{y}_{i1} & \dots & \hat{y}_{ik} & \dots & \hat{y}_{iK}  \\
      \vdots & \ddots & \vdots &\ddots & \vdots\\
     \hat{y}_{n1} & \dots & \hat{y}_{nk} & \dots & \hat{y}_{nK}  \\

    \end{bmatrix}
\]

This represents the final ouptut of the modeling/prediction phase of the workflow

\end{enumerate}

\subsection{Preference Constraint}

The next stage of our approach involves eliciting clients' underlying preferences and ruling out locations that are inconsistent with these preferences. Specifically, we assume that for every client $i = 1, \dots, n$, preferences for each location $k \in \{1, \dots, K\}$ can be expressed by a utility value $u_{ik} \in  \mathbb{R}$. The set of utility values over each $K$ location is arranged in a vector $\vec{u}_{i} = [u_{i1}, u_{i2}, \dots , u_{iK}]$. \\

We further assume that every individual has a utility threshold below which they will find a location unacceptable to live in. We denote this utility threshold $\psi_i$. We denote the subset of acceptable locations for each individual $i$ as $S_i =\{k'\} \subset K$, and define acceptable locations in $\{k'\} \subset K$ as those locations where an individual's utility value is above their utility threshold value $\psi_i$.

\[
S_i = \{k'\} \subset K | u_{ik'} > \psi_i \forall k'
\]

Given every individual has their own utility vectors $\vec{u}_i$ and utility threshold value $\psi_i$, the cardinality of set $S_i$ is different for each $i$. Define $t_i = |S_i|$, the number of acceptable locations for person $i$. We assume $t_i \geq 1$; that is that $S_i$ is non-empty and at least one location is above the utility threshold. \\

We find set $S_i$ acceptable locations for each $i$ with a survey method. As expressed in the main paper, we are agnostic as to which survey device is used, as long as the method allows us to restrict the set of locations to those consistent with $i$'s underlying preferences.

\subsection{Recommendations} 

The final stage of our workflow uses as input the predicted outcome vector $\vec{\hat{y_i}}$ and set of feasible locations $S_i$ to produce a final set of recommendations for individual $i$. This process is formalized as follows: \\

Define $\mu$ as a function with predicted outcome vector $\vec{\hat{y_i}}$ and set of feasible locations $S_i$ as inputs. The function $\mu$ then outputs vector $\hat{y}_{iR}$ which ranks expected outcomes for all $t_i$ locations within feasible set $S_i$.

\[
\mu: (\vec{\hat{y_i}},  S_i) \xrightarrow{}\vec{\hat{y}}_{iR} = [\hat{y}_{ik'_1}, \hat{y}_{ik'_2} ... \hat{y}_{ik'_{t_i}}]
\]

\[
 \mbox{s.t. } \hat{y}_{ik'_1} \geq \hat{y}_{ik'_2} \geq ... \geq \hat{y}_{ik'_{t_i}} \mbox{~and~}k' \in S_i 
\]

Define $z$ as the maximum number of recommendations to present to the user, and $z_i'$ as the minimum between $z$ and $t_i$.
\[z_i' = min(z,t_i)\]

The final interface will recommend the top $z'_i$ locations to user $i$ order. Various formats could be used to present the recommendations. 

$$(k'_1, k'_2, ... , k'_{z'_i})$$

\section{Properties of the Modeling/Prediction Stage}

For individuals denoted by $i$, let $Y_i$ denote observed outcomes, $A_i$ denote their chosen locations, and $X_i$ denote their observed characteristics (which can denote a vector of covariates or a single fully stratifying variable). Further, let $Y_i(a)$ denote the potential outcome for individual $i$ in any location $a \in S_A$, where $S_A$ denotes the set of possible locations. In other words, $Y_i(a)$ represents the outcome that individual $i$ would achieve if that individual had chosen location $a$, and $Y_i = Y_i(A_i)$.\footnote{Note that the definition of the potential outcomes implies the stable unit treatment value assumption (SUTVA).} In the modeling/prediction stage of the decision helper, the goal is to determine the optimal location for each individual as a function of their observed characteristics. In other words, for each stratum $X_i = x$ and at each location $a \in S_A$, the goal is to determine the following quantity of interest:
$$\theta_{a}(x) \equiv E[Y_i(a) | X_i = x]$$
where the expectation (and all expectations presented hereafter) is defined over the distribution of the population of interest (i.e. the population for whom the decision helper is targeted).

The goal is then to use this quantity for all $a \in S_A$ to determine each individual's optimal location(s)---that is, the location(s) for which the quantity is highest, perhaps subject to additional constraints---and then deliver informational nudges to encourage individuals to land in these locations.

However, a key impediment to using $\theta_{a}(x)$ in this ideal manner is that $\theta_{a}(x)$ is not necessarily identifiable with observed data. Instead, what is identified is the following:
$$\theta'_{a}(x) \equiv E[Y_i(a) | X_i = x, A_i = a] = E[Y_i | X_i = x, A_i = a]$$
That is, it is based upon $\theta'_{a}(x)$ (or estimates thereof) that optimal locations will be inferred for each individual, and these inferences may not perfectly match the true optimal locations as defined by $\theta_{a}(x)$.

Therefore, it is useful to characterize the potential bias of $\theta'_{a}(x)$ with respect to $\theta_{a}(x)$ in order to (a) understand the extent to which that bias may result in suboptimal informational nudges and (b) identify concrete actions that can be taken to limit or eliminate the bias. To do so, the following additional quantities are first defined:
$$\theta''_{a}(x) \equiv E[Y_i(a) | X_i = x, A_i \neq a]$$
$$p_{a}(x) \equiv P(A_i = a | X_i = x)$$
In addition, assume that $0 < p_{a}(x)$ for all $a$ and $x$, and note that $\theta_{a}(x) = \theta'_{a}(x) p_{a}(x) + \theta''_{a}(x) (1-p_{a}(x))$. Hence, the bias of $\theta'_{a}(x)$ with respect to $\theta_{a}(x)$ is bounded by the following:
$$B_a(x) \equiv \lim_{p_a(x) \to 0} \left( \theta'_{a}(x) - \theta_{a}(x) \right) = \theta'_{a}(x) - \theta''_{a}(x)$$
$$ = E[Y_i(a) | X_i = x, A_i = a] - E[Y_i(a) | X_i = x, A_i \neq a]$$
This term is a form of selection bias that represents, within a stratum of $x$, the extent to which the mean potential outcome in location $a$ is higher for individuals who actually choose $a$ versus individuals who do not choose $a$. 

If it can be assumed that $Y_i(a) \indep A_i | X_i$, then the selection bias is eliminated (i.e. selection on observables). However, it could be that the potential outcomes are also related to unobserved characteristics of an individual that may also be correlated with location choices. Such unobserved characteristics can be separated into two broad categories. The first, denoted by $U_i$, are unobserved characteristics that are unrelated to any particular location. Examples would include an individual's unmeasured abilities or motivation. The second, denoted by $V_{ai}$, are unobserved characteristics that are unique to a particular location $a$ in question. Examples would include an individual's unknown job offer or social network (e.g. family members) in location $a$.

Taking these unobserved characteristics into account, for any location $a$ let the potential outcome $Y_i(a)$ be modeled as an arbitrary (and arbitrarily complex) function of $X_i$, $U_i$, and $V_{ai}$ as well as an exogenous error term:
$$Y_i(a) = g_a(X_i, U_i, V_{ai}) + \epsilon_i$$
where $E[\epsilon_i | X_i, A_i] = 0$. By extension, we have the following:
$$B_a(x) = E[Y_i(a) | X_i = x, A_i = a] - E[Y_i(a) | X_i = x, A_i \neq a]$$
$$= E[g_a(X_i, U_i, V_{ai}) + \epsilon_i | X_i = x, A_i = a] - E[g_a(X_i, U_i, V_{ai}) + \epsilon_i | X_i = x, A_i \neq a]$$
$$= \int g_a(x,u,v_a) dF_{U_i, V_{ai} | X_i = x, A_i = a}(u, v_a) - \int g_a(x,u,v_a) dF_{U_i, V_{ai} | X_i = x, A_i \neq a}(u, v_a)$$
$$= \int \int g_a(x,u,v_a) dF_{V_{ai} | U_i = u, X_i = x, A_i = a}(v_a) dF_{U_i | X_i = x, A_i = a}(u)$$
$$- \int \int g_a(x,u,v_a) dF_{V_{ai} | U_i = u, X_i = x, A_i \neq a}(v_a) dF_{U_i | X_i = x, A_i \neq a}(u)$$
where $F_{U, V_a | X, A}$ denotes the joint conditional distribution function of $U$ and $V_a$ given $X$ and $A$; $F_{V_a | U, X, A}$ denotes the conditional distribution function of $V_a$ given $U$, $X$, $A$; and $F_{U | X, A}$ denotes the conditional distribution function of $U$ given $X$ and $A$, all in the population of interest.

These results help to highlight what assumptions are required, and what corresponding design decisions could be made, to limit or eliminate this bias. For instance, provided a sufficiently rich set of covariates are observed in $X_i$, the following assumption may hold:
$$U_i \indep A_i | X_i$$
In words, this assumption states that within strata of $X$, individuals are not self-selecting into locations as a function of unobserved non-location-specific characteristics. For instance, for individuals who are identical on $X_i$ (e.g. same age, education, profession, skills, etc.), their variation in unmeasured variables  $U_i$ such as motivation is unrelated to their location choices $A_i$. Under the assumption that $U_i \indep A_i | X_i$, $F_{U | X, A} = F_{U | X}$ and hence the bias term simplifies to:
$$B_a(x) = \int \int g_a(x,u,v_a) \{ dF_{V_{ai} | U_i = u, X_i = x, A_i = a}(v_a) - dF_{V_{ai} | U_i = u, X_i = x, A_i \neq a}(v_a) \} dF_{U_i | X_i = x}(u)$$
In other words, under this assumption, the bias is driven by the difference in the distribution of $V_{ai}$ for individuals who choose $a$ versus do not choose $a$, by joint strata of $X$ and $U$. If we make this assumption in the context of the simulated backtests applied to the Canada Express Entry applicants, then for the resulting estimated gains to be driven purely by bias, this would mean that the average estimated gains among compliers can be accounted for by bias attributed solely to location-specific links or advantages that the individuals who chose any particular Economic Region had over otherwise identical individuals who did not choose that Economic Region. In other words, the individuals who select into a particular location have an average annual employment income advantage at that location of between \$11,000 and \$22,700 (depending on the simulation scenario) due to pre-determined job offers or family/social network ties compared to otherwise identical individuals who chose different locations. 

Another assumption that could be made is that $V_{ai}$ is constant in the population of interest, which could be ensured by design by redefining the population of interest and excluding individuals accordingly, e.g. excluding all individuals likely to have pre-determined job offers.\footnote{Note that excluding those with job offers from the training data set would have meant excluding a significant proportion of immigrants who came through Express Entry in the first 2 years (2015-2016). However, this limitation becomes less salient as the share of admissions with job offers has declined considerably in recent years with the reduction in number of points for arranged employment in the CRS. For example, in 2017-2019 only about 10\% of invited candidates had a job offer or arranged employment.} Under the assumption that $V_{ai} = \tilde{v}_a$ for all individuals in the population of interest, the bias term simplifies to the following:
$$B_a(x) = \int g_a(x,u,\tilde{v}_a) dF_{U_i | X_i = x, A_i = a}(u) - \int g_a(x,u,\tilde{v}_a) dF_{U_i | X_i = x, A_i \neq a}(u)$$
If the previous assumption that $U_i \indep A_i | X_i$ is added back in, the bias is completely eliminated:
$$B_a(x) = \int g_a(x,u,\tilde{v}_a) \{ dF_{U_i | X_i = x}(u) - dF_{U_i | X_i = x}(u) \} = 0$$

\section{Application of Decision Helper Workflow: Canada Express Entry}

The empirical application of our proposed decision helper workflow analyzed data from Canada's Express Entry system. While we provide an overview of this method in the body of our paper, here we provide  additional methodological details on how we implement the \textbf{Model/Predictions} and \textbf{Preference Constraint} portion of our workflow.

\subsection{Data Sources}

We merged three datasets at the Federal Research Data Center in Ottawa to conduct our analysis: 

\begin{itemize}
\item IMDB Integrated Permanent and Non-permanent Resident File (PNRF) 1980-2018 (2019 release)
\item IMDB Tax Year Files (t1ff) 2013-2017 (2019 release) 
\item Express Entry File (2018 release, case level data)
\end{itemize}

In addition, we leveraged several additional datasets to provide supplementary information on population levels, unemployment rates, and price indices by geographic region:

\begin{itemize}
\item Canada Labour Force Survey (LFS): A monthly survey providing data on the labour market, including estimates of employment and demographics of the working population. Estimates are available at different levels of geographic aggregation, including Economic Region.
\item Canadian Rental Housing Index: Public index compiled by the BC Non-Profit Housing Association, based on the 2016 Census. The index reports the average rental price for a single-family apartment, by Census Subdivision (CSD). 
\end{itemize}

A full list of considered variables is found in Table \ref{tab:varnames}. Note that all analyses was conducted in the Federal Research Data Centre in Ottawa and all data output presented here was approved for release.

\subsubsection{Administrative Data}

 Although the population of interest consists of immigrants entering through Canada's Express Entry system, limited data on this relatively new initiative required us to expand our training data to similar economic immigrants entering Canada. Specifically, in addition to including all Express Entry clients entering between 2015 and 2016 who filed a tax return, we expand our training set to include Non-Express Entry clients who entered between 2012 and 2016 and filed a tax return under four admission categories that would be managed by the Express Entry system: Federal Skilled Workers (A1111), Skilled Trades (A1120), Canadian Experience (A1130), and Provincial Nominees (A1300). \\

We restrict our training set by removing:

    \begin{itemize}
        \item Individuals who were selected by Quebec or landed in Quebec, given that Express Entry does not apply to this province.
         \item Individuals whose yearly income in the year after arrival exceeded the 99th percentile, to avoid overfitting to outliers.
         \item Accompanying children (LANDING\_AGE $>$ 18)
         \item Individuals who died in the year of arrival or in the following year
    \end{itemize}

This set of training data corresponds to matrix \textbf{A} in the decision helper helper workflow.

\subsubsection{Client Data}

The client dataset consists of the Express Entry cohort who entered prior to 2017 (the final year available in the outcome data), along with their associated characteristics from the PNRF file. This represents the group of economic immigrants we consider in all our simulation results. A set of descriptive statistics for this subgroup is found in Table \ref{tab:descrstats}.

This set of prediction data corresponds to \textbf{C} in the general decision helper helper workflow.

\subsection{Modeling Decisions} Our workflow allows for a wide variety of potential models to be used in predicting outcomes. In this section, we describe the particular modeling decisions we made in the context of analyzing Canadian immigration data.

\subsubsection{Models}

We used a supervised machine learning framework to fit and train our models. We use this class of models due to their ability to both flexibly fit the training data while retaining high out-of-sample accuracy with proper model tuning. While any number of supervised machine learning methods might be applicable, we chose to use gradient boosting machines due to their ability to automatically engage in feature selection and discover complex interactions between covariates. \\

We implement the modeling stage on a location-by-location basis. Specifically, for each economic region, we first subset the training data to those individuals who originally landed in that location, and fit the supervised learning model using individuals' background characteristics to predict their employment earnings. 
We model synergies using stochastic gradient boosted trees, which we run with a customized script using the gradient boosting machine (gbm) package within R. \\

In our implementation of gradient boosted trees, we used 10-fold cross-validation within the training data to select tuning parameter values, including the interaction depth (the maximum nodes per tree), bag fraction (the proportion of the training set considered at each tree expansion), learning rate (the size of each incremental step in the algorithm), and number of boosting iterations (number of trees considered). \\

To determine the best model, we first fix an interaction depth, bag fraction, and learning rate. For this fixed set of parameters, we then fit models over a sequence of boosting iterations (normally, 1 to 1,000 trees). For each model, we calculate the cross-validation root mean square error (RMSE), and choose the  model which minimizes this error. To avoid potentially choosing a local minimum, if the best model is within 100 trees of the maximum number of trees we consider, we re-run this process by increasing the maximum considered trees by 500. We repeat this process as many times as necessary, and record the final tree count and RMSE for the fixed interaction depth, bag fraction, and learning rate. \\

We repeat the above process tuning over different values of interaction depth, bag fraction, and learning rate. Finally, we pick the model with the set of parameters with the lowest cross-validation RMSE for each separate location model. The set of parameters we consider are:

\begin{itemize}
    \item Interaction Depth: 5-7
    \item Learning Rate: .1 and .01
    \item Bag Fraction: .5-.8 by .15
\end{itemize}

The set of final models, one for each location, correspond to the set of \textbf{L} models in the general workflow.\\

\begin{figure}
    \centering
    \includegraphics[width=1\linewidth]{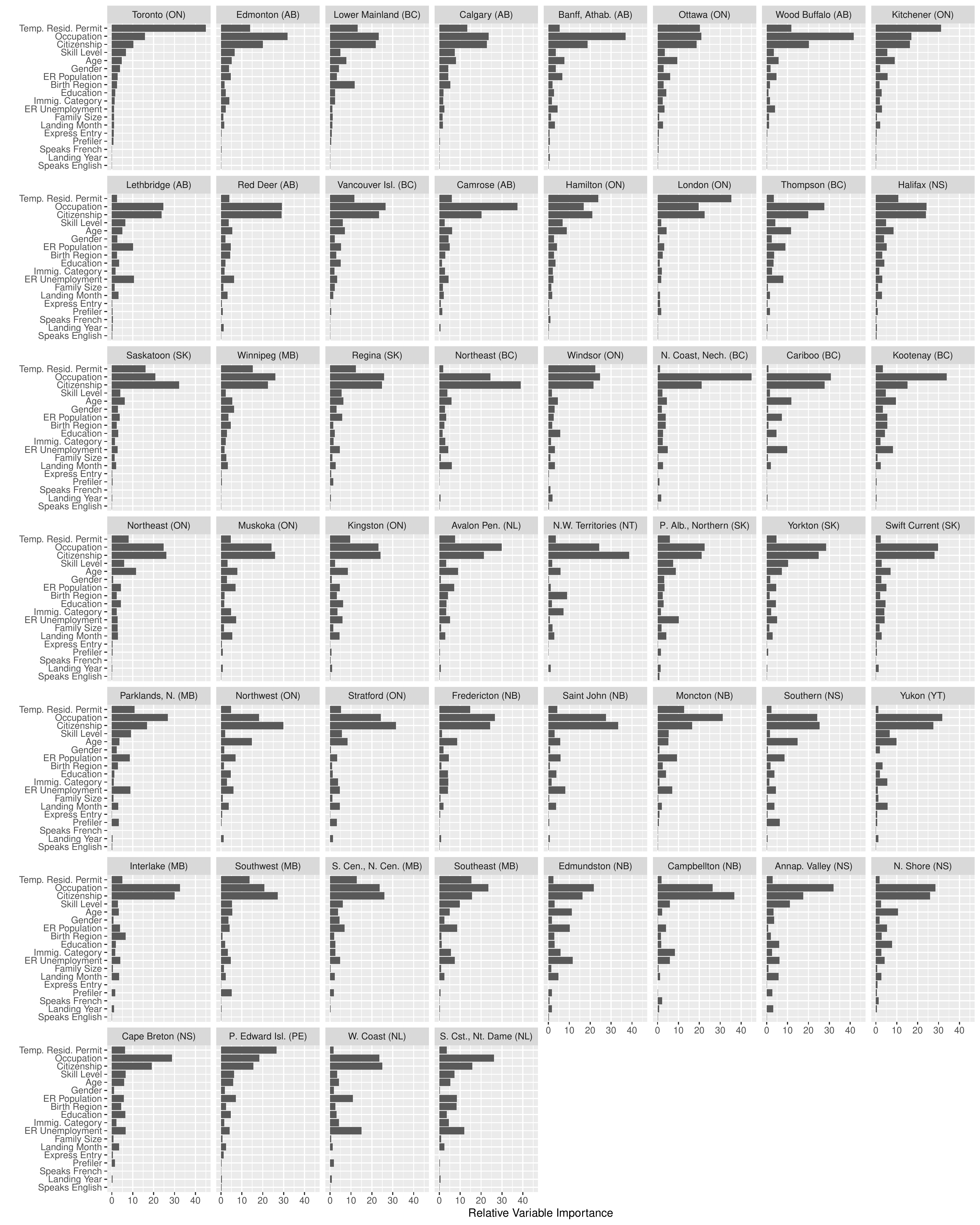}
    \caption{Variable Importance Statistics for Tuned Location Models (Principal Applicants)}
    \label{fig:s0}
\end{figure}

In order to investigate which covariates are the most predictive of income, we calculate a variable importance measure for each predictor in every separate tuned location model  (see summary.gbm in the gbm package for details on how this statistic is calculated). We present these variable importance measures in Figure \ref{fig:s0}.\\

This figure demonstrates one of the advantages of fitting each location model separately -- in each model, the importance measures of each covariate differs, demonstrating how the set of characteristics that lead to better or worse economic outcomes vary between ERs. Some overall trends emerge, with occupation and citizenship in the top most influential covariates in every model. Whether or not a client had a previous temporary residence permit is also a highly influential predictor in certain Economic regions, especially in Toronto. We further note certain variables  have little influence on predicted income across each model, such as the language (French and English) indicators and landing year.

\subsubsection{Predicted Outcomes}

We then apply each fitted model to the prediction set to estimate the income for new Express Entry clients should they select the economic region in question. For the prediction set, we remove any Express Entry client with the provincial nomination (A1300) category, as these clients do not have flexibility in choosing the initial landing location. This process is performed separately and independently for each location, which yields a vector of predicted income across possible economic regions for each individual within the prediction set. The final result is a matrix of predicted annual income with rows representing individual Express Entry clients and columns representing economic regions,  corresponding to \textbf{M} in the general workflow. \\

\begin{figure}
    \centering
    \includegraphics[width=.95\linewidth]{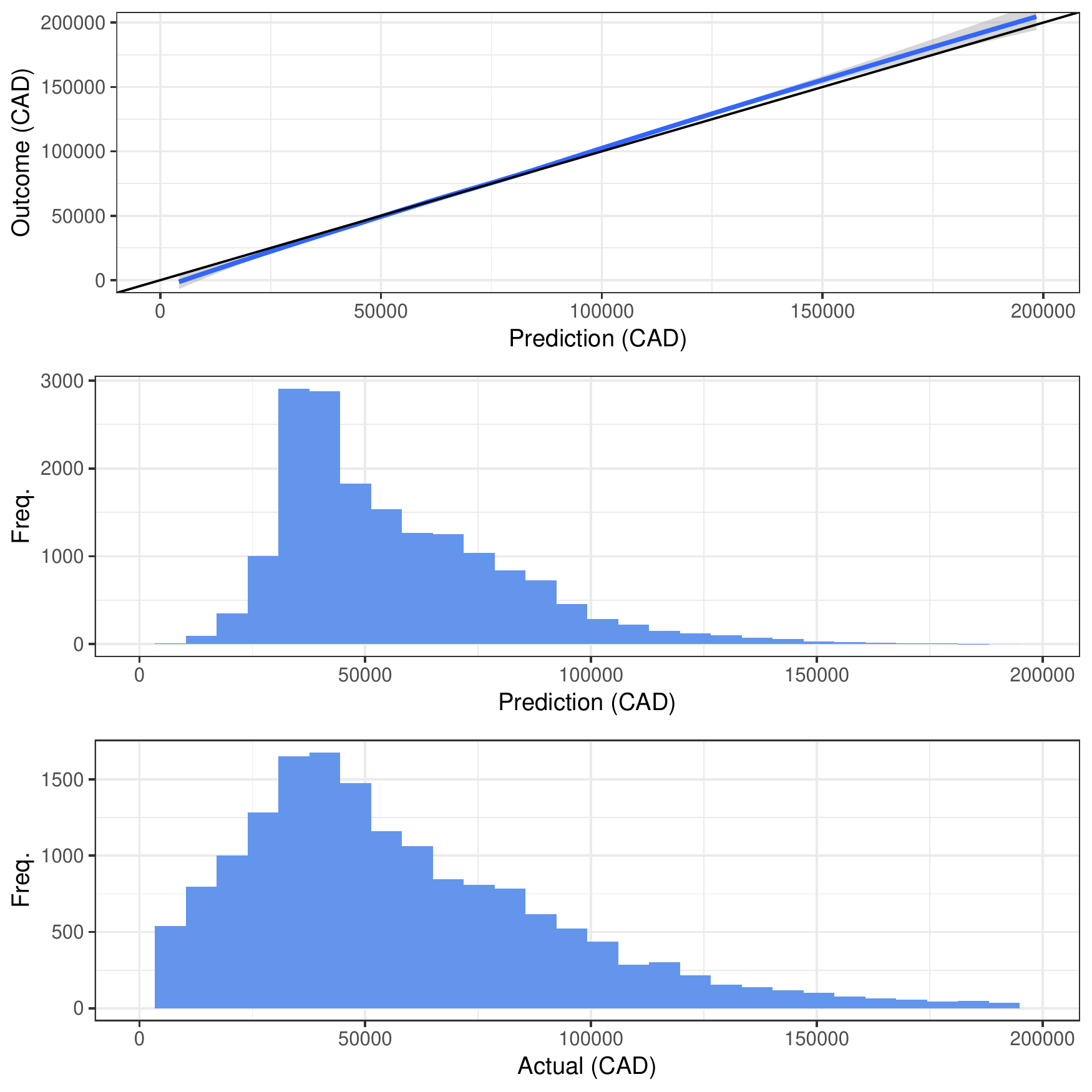}
    \caption{Calibration Plot: Model Fit for Express Entry Clients}
    \label{fig:s1}
\end{figure}

In order to asses model fit, we compare predicted income within the principal applicants' actual location to their observed  income in that location in the top panel of Figure \ref{fig:s1}. The bottom two panels show the histogram of predicted incomes and actual incomes respectively. Overall, predictions are well calibrated, albeit slightly more conservative than observed income at the tails of the distribution. \\

In our implementation, the cross-validated R$^{2}$ for the tuned PA model is 0.54. Within the context of incomes -- which are highly skewed and difficult to predict -- this is relatively high. This represents a substantial improvement over the R$^{2}$ for an analogous linear regression model using cross-validation (0.34). The RMSE for the tuned model is 29,486, as compared to an observed mean income of 58,000 and a standard deviation of 41,600.

\subsection{Approximating Locational Preferences}

While a prospective use case would use a survey restrict locations to a set that align with an individual's underlying preferences, we were unable to engage in this exercise in our backtests. Therefore, we use the administrative data and existing migration patterns to estimate preferences. We explain the details of this methodology in this section.

\subsubsection{Underlying Preferences}

Upon entering Canada, clients plausibly have a series of idiosyncratic locational preferences related to geographic location, climate, local demographics, labor markets, and other potential factors. We approximated individual locational preferences by examining how Express Entry clients with different background characteristics varied in terms of their original landing locations. Using the landing Economic Region as the dependent variable, we fit a multinomial logit model on Express Entry clients and a random subset of 20\% of the non-Express Entry economic immigrants. Given that these preferences are proxies, we use a coarse set of covariates including education, birth region, age, immigration category, case size, and indicators for work and study permits.  These predictions approximate the vector of utilities \textbf{u} in the workflow.

\subsubsection{Restricted Set of Locations}

After obtaining predictions for each Express Entry case, we rank order locations by predicted preferences, randomly breaking ties. The resulting ranks are used in conjunction with the parameter $\phi$ (the number of acceptable locations we consider) to determine the initial set of locations considered when selecting the locations with the top optimal income. This set of $\phi$ acceptable locations represents subset $\mathbf{S_i}$ in the workflow.

\section{Robustness Checks}

Below we discuss several robustness checks to our study. In particular, we outline the impact of our backtest when considering 1) cost-of-living adjustments, 2) maximizing principal applicant plus spouse income, and 3) alternative simulation specifications.

\subsection{Cost-of-living Adjustments}
An important consideration influencing relative quality of life across landing locations relates to living costs. To take this factor into account during the recommendation process, we ran alternate simulations where we define outcomes as total income less estimated yearly rental costs of a two bedroom apartment. To our knowledge, rental prices are the most granular cost index currently available across small  geographic regions. The estimates we pull are from the Canadian Rental Housing Index, a public index compiled by the BC Non-Profit Housing Association and based on the 2016 Census. The index reports the average rental price for a single-family apartment, by Census Subdivision (CSD). \\

\begin{figure}
    \centering
    \includegraphics[width=.95\linewidth]{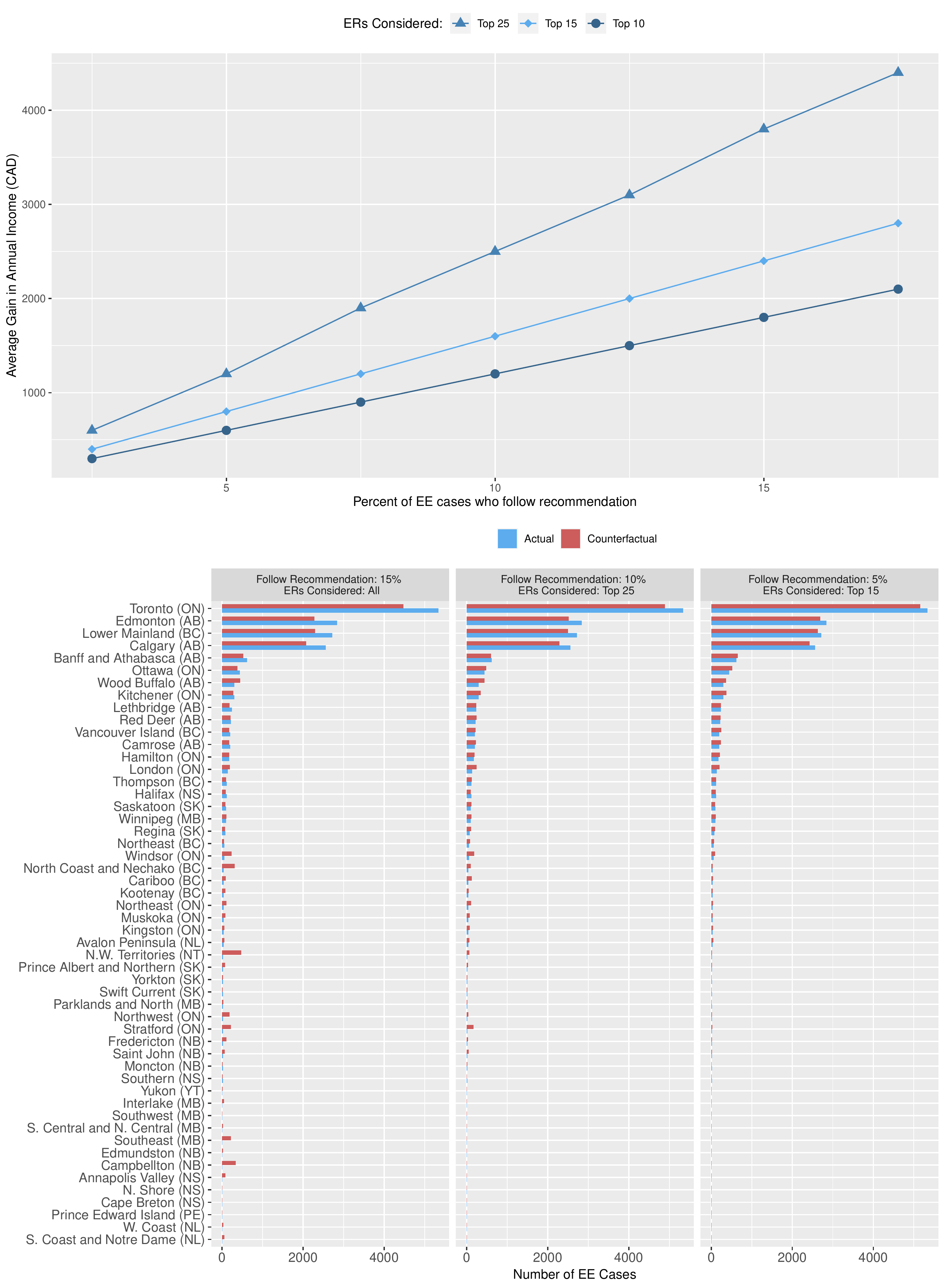}
    \caption{Estimated Average Income Gains and Shifts in Arrival Locations with CPI Adjustments}
    \label{fig:s2}
\end{figure}

In Figure \ref{fig:s2}, we replicate the results in our main paper, demonstrating average gains in employment under various simulation parameters in the top panel and visualizing expected movement patterns in the bottom panel.

\subsection{Principal Applicant and Spouse Model}

\begin{figure}
    \centering
    \includegraphics[width=.95\linewidth]{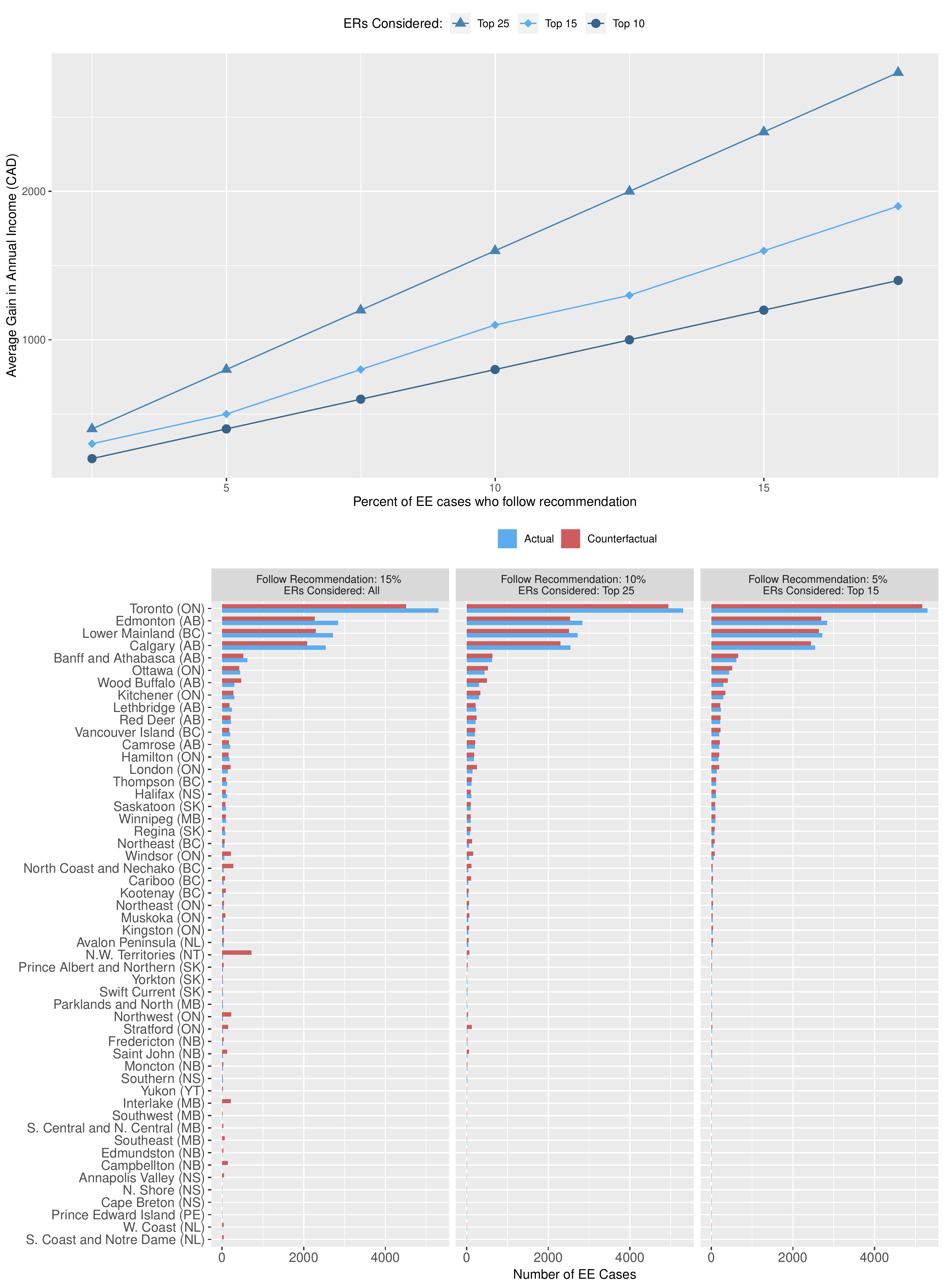}
    \caption{Estimated Average Income Gains and Shifts in Arrival Locations for Principal Applicant and Spouse  Model}
    \label{fig:s3}
\end{figure}

While our main paper reports our findings for  principal applicants only, we additionally fit a set of models that consider both principal applicants and their accompanying partners. As a simplifying assumption, we assume that both individuals in a case have similar (joint) locational preferences, and derive these preferences from a PA-only model. The income predictions, however, take into account the joint income of the PA and partner divided by the number of adults in the family unit (average family income). We then estimate the models assuming a family unit will move jointly. In Figure \ref{fig:s3}, we replicate the results in the main paper using this approach, with similar results.

\subsection{Alternative Simulation Specifications}

Our simulations vary two parameters: the number of acceptable locations considered ($\phi$) and the compliance rate ($\pi$). In this section, we consider the impact of further varying these parameters on our core results.

\subsubsection{No Locational Preferences}

\begin{figure}
    \centering
    \includegraphics[width=.95\linewidth]{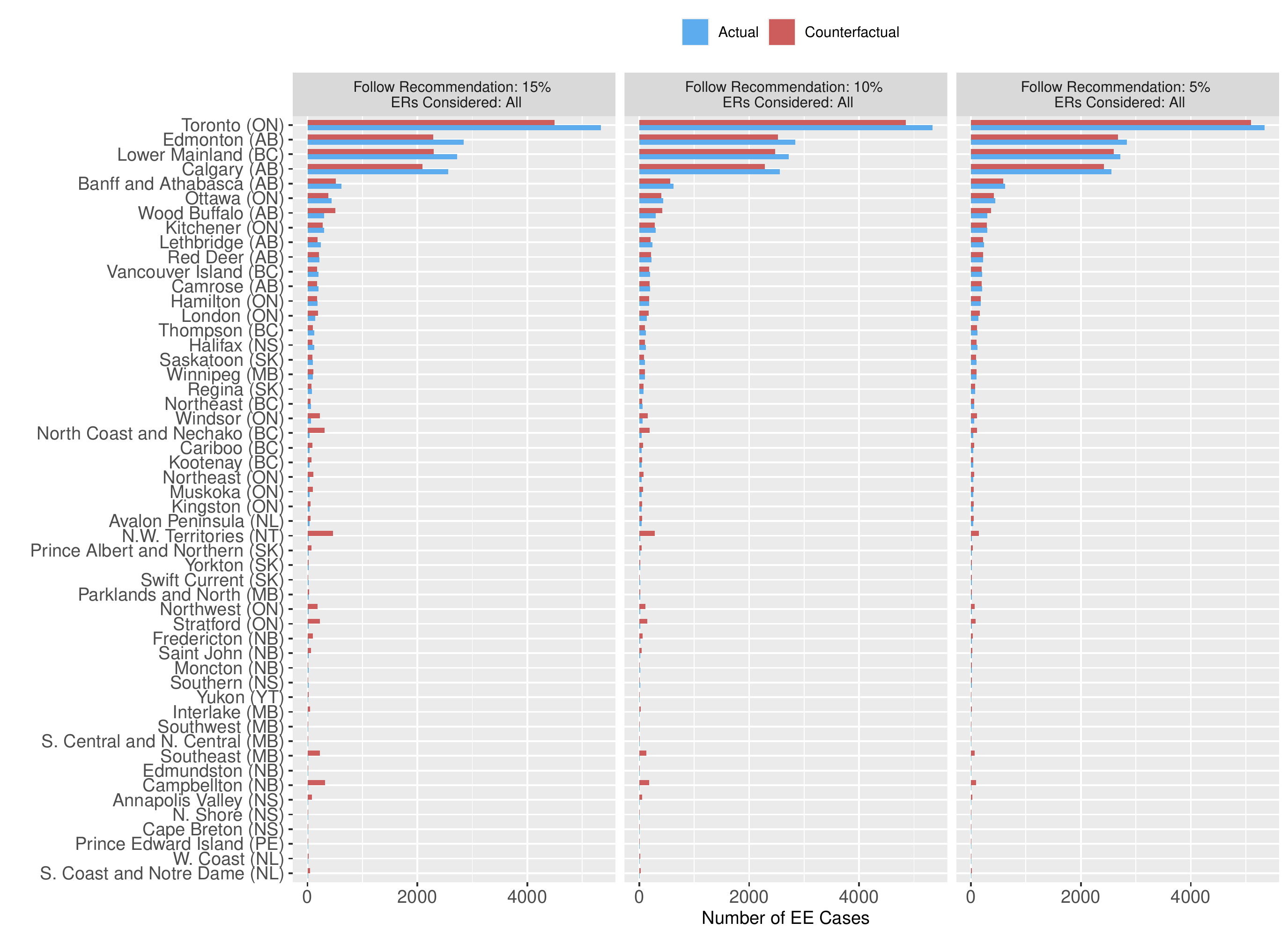}
    \caption{Movement Under Various Simulation Parameters}
    \label{fig:s4}
\end{figure}

In our main analysis, we infer regional preferences by analyzing existing residential patterns and then using these estimated preferences to restrict the choice set in our simulations. However, expected income gains are maximized when no locational preferences are taken into account. In Figure \ref{fig:s4}, we show simulated movement patterns with no locational preferences under different compliance rates. Relative to the case presented in the main paper, the results similarly suggest that the majority of outflow is from the four largest locations, but display increased recommendations to smaller locations. 

\subsubsection{Constant Compliance Rate}

In the body of the paper, we present results where we vary the compliance rate $\pi$ as a function of income. Specifically, we specify an upper bound $\pi_{max}$ to the individuals with the lowest actual income before linearly interpolating to the value $\pi = 0$ across the entire income distribution. Thus, each individual receives a heterogeneous compliance parameter $\pi_i$, and the average compliance rate in a particular simulation run is reported as $\pi_{max}$/2. \\

\begin{figure}
    \centering
    \includegraphics[width=.8\linewidth]{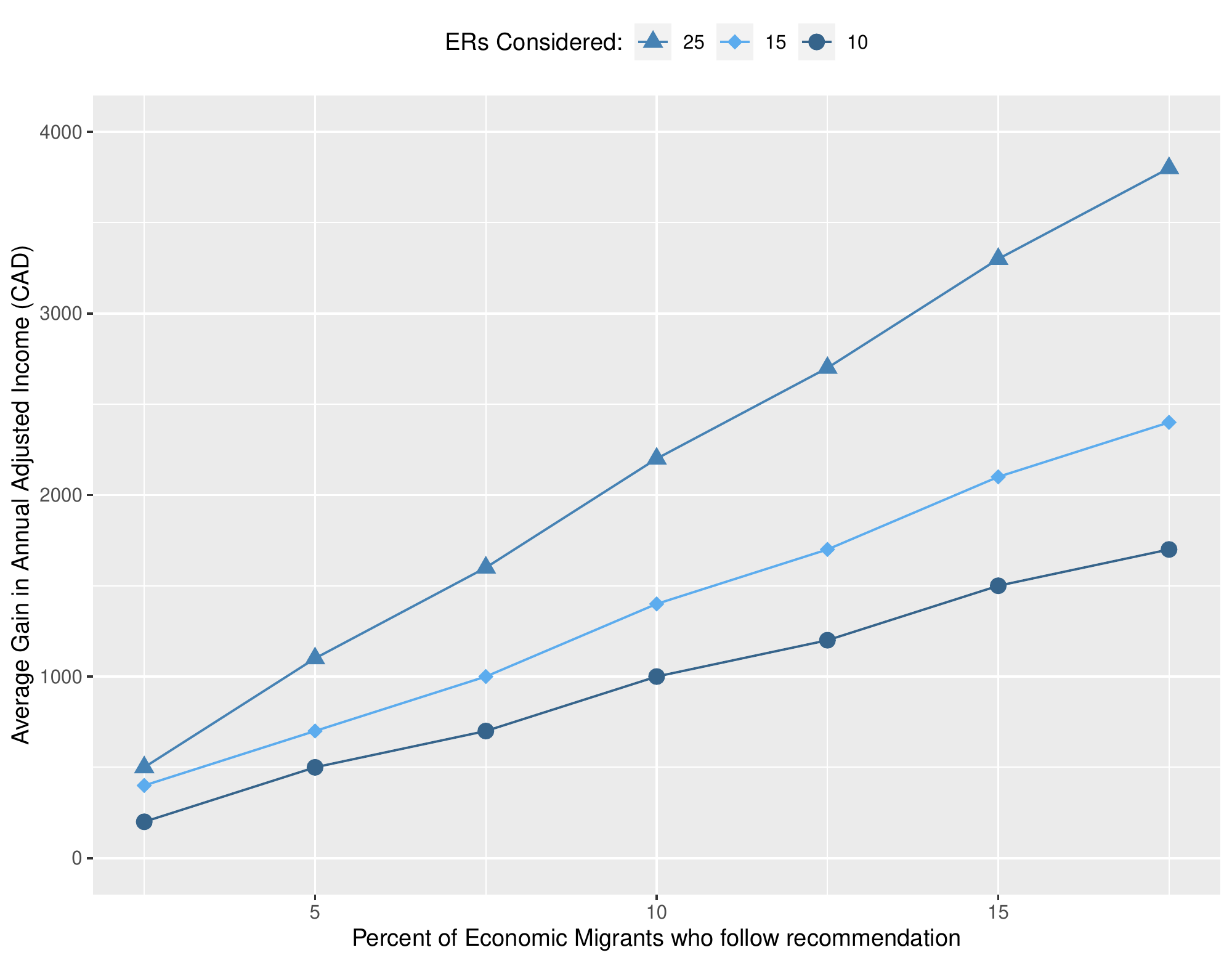}
    \caption{Constant Compliance Rate: Estimated Average Income Gains. N=17,640}
    \label{fig:s5}
\end{figure}

To ensure that results are not driven by this modeling decision, we repeat our simulations with a constant compliance rate in an individual simulation run. In each of these tests, we set a single $\pi$ that represents each individuals' likelihood of complying, which is constant across income. We present these results in Figure \ref{fig:s5}, which reveals very similar potential average gains in income across each simulation.

\subsubsection{Removing Economic Regions}

In order to evaluate whether the gains we report in the main analysis of our paper are being driven by a specific subset of ERs, we rerun the simulations exactly as described in the body of the paper but remove from consideration certain subsets of landing locations. That is, if an individual `complies' with probability $\pi_i$, we limit the set of locations they can potentially move to in the simulation.

We begin by specifying three alternative models, in each case excluding a subset of ERs that could potentially drive our results. In the first alternative model, we do not allow individuals to move to the largest ERs, those with a population greater than 1,500,000 according to the 2016 census. In the second, we extend this to include large and growing ERs, defined as all ERs with a population greater than 1,000,000 and a growth rate in population from the 2011 to 2016 census above 5\%. In the third, we remove the smallest ERs -- those with a population less than 100,000 in the 2016 census. A full list of removed ERs in each specification is listed in the Table \ref{tab:ER_Remove} below.

\begin{figure}
    \centering
    \includegraphics[width=.95\linewidth]{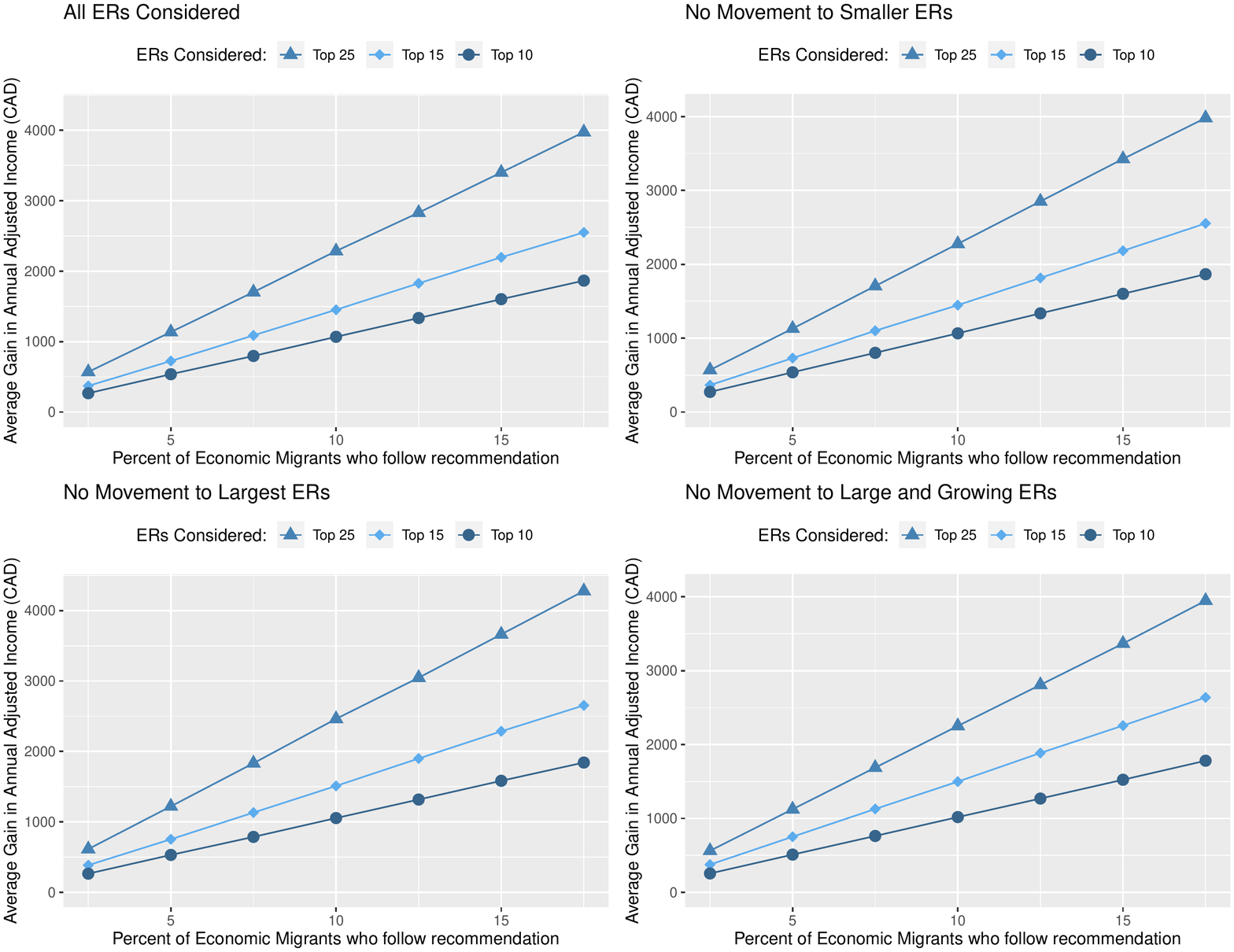}
    \caption{Removing Subsets of ERs. N=17,640}
    \label{fig:s6}
\end{figure}

The results of these alternative model runs are found in Figure \ref{fig:s6}. Overall, these plots demonstrate the gains we find in our main analysis are not driven by one of the ER subsets we define above. In the top-left panel, we presents results from a simulation considering ``All ERs,'' effectively replicating our main results in the body of our paper. In each of the three alternative specifications, we see that average gains across each compliance and preference parameter do not substantially differ from this baseline.

\begin{figure}
    \centering
    \includegraphics[width=.95\linewidth]{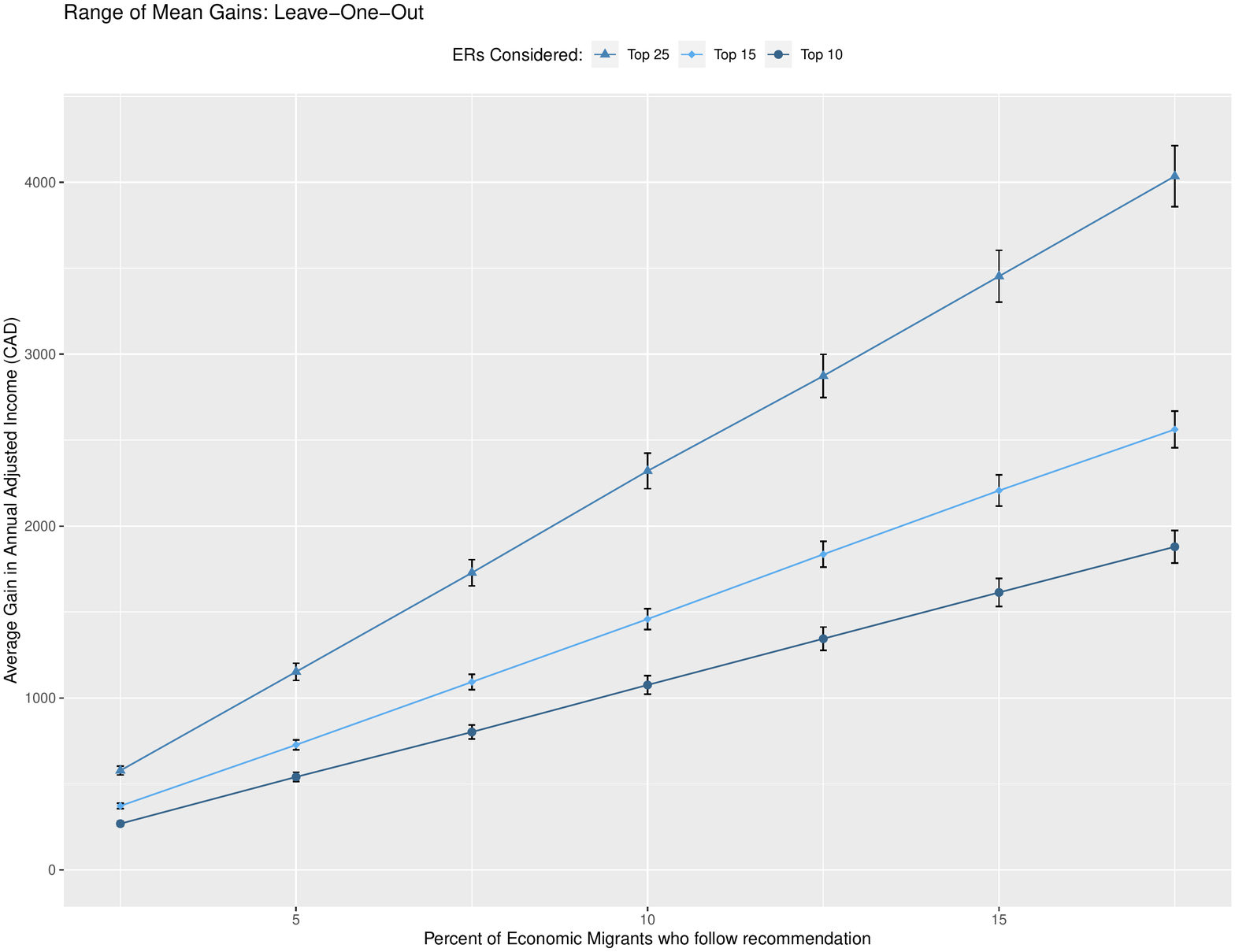}
    \caption{Average Gains Across 52 Leave-One-Out Simulations. N=17,640}
    \label{fig:s7}
\end{figure}

Another way we check against the impact of a single ER on our core results is by running a ``leave-one-out'' robustness check. In this test, we run a series of 52 simulations, in each case dropping one of the 52 total ERs from consideration. Other than removing this single ER from the choice set, the simulations are run exactly as described in the body of the paper. For simulation, we calculate the mean gain in annual income. We present the average of these gains and the 95\% confidence interval across the 52 simulations  in Figure \ref{fig:s7}. We again see little change to our core results, demonstrating no single ER drives the average gain in income in our simulations.

\clearpage

\section{Figures}

{}

{}

\newpage
{}

\newpage
{}

\newpage
{}

\newpage
{}

\newpage
{}

\newpage
{}

\newpage
\clearpage
\section{Tables}

\clearpage

\begin{table}[!h] \small
\begin{tabularx}{\textwidth}{llX}
Original Variable Name & Source & Description \\
\hline
COUNTRY\_BIRTH$^{1,4}$&PNRF\\
COUNTRY\_CITIZENSHIP$^{1,4}$&PNRF\\
CSQ\_IND$^{2}$&PNRF & Quebec program indicator\\
DEATH\_INDICATOR$^{2}$&PNRF\\
DESTINATION\_ER$^{1,2,3}$&PNRF& Intended Economic Region (ER) of landing\\
EDUCATION\_QUALIFICATION$^{1}$&PNRF\\
EXPRESS\_ENTRY\_IND$^{1,2}$&PNRF&Express Entry flag\\
FAMILY\_STATUS$^{1,2}$&PNRF\\
GENDER$^{1}$&PNRF\\
IMMIGRATION\_CATEGORY\_CENSUS$^{1,2}$&PNRF& Admission category\\
LANDING\_AGE$^{1}$ & PNRF \\
LANDING\_MONTH$^{1}$&PNRF\\
LANDING\_YEAR$^{1,2}$&PNRF\\
LEVEL\_OF\_EDUCATION$^{1}$&PNRF\\
NOC3\_CD11$^{1}$&PNRF & 3-digit expected occupation code\\
NUMBER\_STUDY\_PERMITS$^{1}$&PNRF\\
NUMBER\_WORK\_PERMITS$^{1}$&PNRF\\
OFFICIAL\_LANGUAGE$^{1}$&PNRF\\
SKILL\_LEVEL\_CD11$^{1}$&PNRF\\
EI\_\_\_I$^{1}$&Tax&Individual employment income (excluding self-employment)\\
PREFILER\_IND$^{1}$&Tax&Whether an individual filed a return on a TR permit\\
POPULATION\_ER$^{1}$ & LFS & Quarterly population \\ 
PRICE\_INDEX$^{3}$ & External & Average yearly rental price \\ 
UNEMPLOYMENT\_ER$^{1}$ & LFS & Quarterly unemployment \\ 
\hline
\\
\textsuperscript{1} = Variable used to train machine learning models \\
\textsuperscript{2} = Variable used to subset the data \\ 
\textsuperscript{3} = Variable used to adjust final predictions \\ \\
\textsuperscript{4} = Variable aggregated to the continent level for modeling \\ \\
\end{tabularx}
\caption{Variable Names}
\label{tab:varnames}
\end{table}

\begin{table}[!h]
\centering\begingroup\fontsize{5}{7}\selectfont

\begin{tabular}{lrrrr|lllll}
\toprule
 & mean & sd & min & max &  & mean & sd & min & max\\
\midrule
 Annual Income per Head (CAD) & 49900.00 & 33600.00 & 0 & 246600 & English: No & 0.02 & 0.14 & 0 & 1\\
Age & 33.15 & 5.99 & 22 & 65 & English: Yes & 0.98 & 0.14 & 0 & 1\\
 Birth Region: The Americas & 0.09 & 0.29 & 0 & 1 & French: No & 0.97 & 0.16 & 0 & 1\\
Birth Region: Europe & 0.24 & 0.43 & 0 & 1 & French: Yes & 0.03 & 0.16 & 0 & 1\\
 Birth Region: Africa & 0.06 & 0.23 & 0 & 1 & Prefiler: No & 0.13 & 0.34 & 0 & 1\\
Birth Region: Asia & 0.59 & 0.49 & 0 & 1 & Prefiler: Yes & 0.87 & 0.34 & 0 & 1\\
 Birth Region: Oceania & 0.02 & 0.14 & 0 & 1 & TR: No TR & 0.12 & 0.32 & 0 & 1\\
Citizenship: United States & 0.03 & 0.17 & 0 & 1 & TR: Study & 0.01 & 0.09 & 0 & 1\\
 Citizenship: Mexico & 0.02 & 0.13 & 0 & 1 & TR: Study+Work & 0.31 & 0.46 & 0 & 1\\
Citizenship: Jamaica & 0.01 & 0.11 & 0 & 1 & TR: Work & 0.57 & 0.5 & 0 & 1\\
 Citizenship: Brazil & 0.01 & 0.11 & 0 & 1 &  &  &  &  & \\
Citizenship: France & 0.03 & 0.16 & 0 & 1 &  &  &  &  & \\
 Citizenship: Germany & 0.01 & 0.11 & 0 & 1 &  &  &  &  & \\
Citizenship: Poland & 0.01 & 0.10 & 0 & 1 &  &  &  &  & \\
 Citizenship: Russia & 0.01 & 0.10 & 0 & 1 &  &  &  &  & \\
Citizenship: Ukraine & 0.01 & 0.12 & 0 & 1 &  &  &  &  & \\
 Citizenship: Ireland & 0.05 & 0.21 & 0 & 1 &  &  &  &  & \\
Citizenship: United Kingdom & 0.06 & 0.24 & 0 & 1 &  &  &  &  & \\
 Citizenship: Nigeria & 0.02 & 0.15 & 0 & 1 &  &  &  &  & \\
Citizenship: South Africa & 0.01 & 0.10 & 0 & 1 &  &  &  &  & \\
 Citizenship: Iran & 0.01 & 0.09 & 0 & 1 &  &  &  &  & \\
Citizenship: China & 0.04 & 0.20 & 0 & 1 &  &  &  &  & \\
 Citizenship: South Korea & 0.03 & 0.16 & 0 & 1 &  &  &  &  & \\
Citizenship: Philippines & 0.14 & 0.35 & 0 & 1 &  &  &  &  & \\
 Citizenship: India & 0.28 & 0.45 & 0 & 1 &  &  &  &  & \\
Citizenship: Pakistan & 0.02 & 0.12 & 0 & 1 &  &  &  &  & \\
 Citizenship: Australia & 0.02 & 0.14 & 0 & 1 &  &  &  &  & \\
Citizenship: Other & 0.18 & 0.38 & 0 & 1 &  &  &  &  & \\
 Education: Less than BA & 0.41 & 0.49 & 0 & 1 &  &  &  &  & \\
Education: BA & 0.27 & 0.44 & 0 & 1 &  &  &  &  & \\
 Education: MA & 0.28 & 0.45 & 0 & 1 &  &  &  &  & \\
Education: PhD & 0.03 & 0.18 & 0 & 1 &  &  &  &  & \\
 Male & 0.67 & 0.47 & 0 & 1 &  &  &  &  & \\
Female & 0.33 & 0.47 & 0 & 1 &  &  &  &  & \\
 Unit Size & 1.43 & 0.50 & 1 & 3 &  &  &  &  & \\
Landing Year: 2015 & 0.29 & 0.45 & 0 & 1 &  &  &  &  & \\
 Landing Year: 2016 & 0.71 & 0.45 & 0 & 1 &  &  &  &  & \\
Landing Month (1-12) & 7.40 & 3.40 & 1 & 12 &  &  &  &  & \\
 Category: Skilled Worker program & 0.46 & 0.50 & 0 & 1 &  &  &  &  & \\
Category: Skilled Trades program & 0.09 & 0.29 & 0 & 1 &  &  &  &  & \\
 Category: Canadian Experience Class & 0.45 & 0.50 & 0 & 1 &  &  &  &  & \\
Industry: ArtCultureSport & 0.04 & 0.20 & 0 & 1 &  &  &  &  & \\
 Industry: Computer & 0.20 & 0.40 & 0 & 1 &  &  &  &  & \\
Industry: Education\_Law\_Govt & 0.07 & 0.25 & 0 & 1 &  &  &  &  & \\
 Industry: Extraction & 0.00 & 0.03 & 0 & 1 &  &  &  &  & \\
Industry: Finance & 0.09 & 0.29 & 0 & 1 &  &  &  &  & \\
 Industry: FoodTourism & 0.12 & 0.33 & 0 & 1 &  &  &  &  & \\
Industry: Health & 0.04 & 0.21 & 0 & 1 &  &  &  &  & \\
 Industry: Management\_Misc & 0.03 & 0.16 & 0 & 1 &  &  &  &  & \\
Industry: ManualTrades & 0.10 & 0.31 & 0 & 1 &  &  &  &  & \\
 Industry: Manufacturing & 0.01 & 0.08 & 0 & 1 &  &  &  &  & \\
Industry: NatResources\_9980 & 0.01 & 0.10 & 0 & 1 &  &  &  &  & \\
 Industry: No\_Info & 0.01 & 0.08 & 0 & 1 &  &  &  &  & \\
Industry: Sales & 0.05 & 0.22 & 0 & 1 &  &  &  &  & \\
 Industry: Services & 0.18 & 0.38 & 0 & 1 &  &  &  &  & \\
Industry: SocialServices & 0.01 & 0.12 & 0 & 1 &  &  &  &  & \\
 Industry: Technical & 0.03 & 0.18 & 0 & 1 &  &  &  &  & \\
Skill Level: Managerial & 0.10 & 0.30 & 0 & 1 &  &  &  &  & \\
 Skill Level: Professionals & 0.35 & 0.48 & 0 & 1 &  &  &  &  & \\
Skill Level: Skilled and Technical & 0.55 & 0.50 & 0 & 1 &  &  &  &  & \\
\bottomrule
\end{tabular}
\caption{Descriptive Statistics: Express Entry Principal Applicants}
\label{tab:descrstats}
\endgroup{}
\end{table}

\begin{table}[ht]
\centering
\begin{tabular}{rrlrrr}
  \hline
ER\_CODE & Name & Large & Rapidly Growing & Small \\ 
  \hline
 3530 & Toronto & X & X &  \\ 
5920 & Lower Mainland & X & X &  \\ 
4830 & Calgary &  & X &  \\ 
4860 & Edmonton &  & X &  \\ 
3540 & Kitchener &  & X &  \\ 
4660 & Interlake &  &  & X \\ 
4680 & Parklands and North &  &  & X \\ 
4840 & Banff and Athabasca &  &  & X \\ 
4740 & Yorkton &  &  & X \\ 
1350 & Edmundston &  &  & X \\ 
5980 & N.E. (B.C.) &  &  & X \\ 
4620 & S. Central and N. Central &  &  & X \\ 
5960 & North Coast and Nechako &  &  & X \\ 
4640 & S. Central and N. Central &  &  & X \\ 
6110 & N.W. Territories &  &  & X \\ 
4670 & Parklands and North &  &  & X \\ 
5970 & North Coast and Nechako &  &  & X \\ 
4760 & Prince Albert and Northern &  &  & X \\ 
1020 & S. Coast and Notre Dame &  &  & X \\ 
6010 & Yukon &  &  & X \\ 
   \hline
\end{tabular}
\caption{Removing Economic Regions Robustness Check}
\label{tab:ER_Remove}

\end{table}

\end{document}